\documentclass[11pt,a4paper]{article}
\pdfoutput=1
\usepackage{jheppub}
\usepackage{slashed}

\makeatletter
\def\@fpheader{\relax}
\makeatother

\usepackage[czech,english]{babel}
\usepackage{graphicx}

\usepackage{amsmath,amsfonts,amssymb}
\usepackage{url}
\DeclareMathOperator{\MyProd}{\scalebox{1.4}{$\mathrm{I\kern-0.2ex I}$}}

\preprint{LCTP-21-03}

\title{Microscopic Entropy of AdS$_3$ Black Holes Revisited}

\author[a]{Finn Larsen}

\emailAdd{larsenf@umich.edu}

\author[a]{and Siyul Lee}

\emailAdd{siyullee@umich.edu}

\affiliation[a]{Leinweber Center for Theoretical Physics, University of Michigan, Ann Arbor, MI 48109, U.S.A.}

\abstract{
We revisit the microscopic description of AdS$_3$ black holes in light of recent 
progress on their higher dimensional analogues. The grand canonical partition function that follows from 
the AdS$_3$/CFT$_2$ correspondence describes BPS and nearBPS black hole thermodynamics.
We formulate an entropy extremization principle that accounts for both the black hole entropy and a constraint on its charges,
in close analogy with asymptotically AdS black holes in higher dimensions. 
We are led to interpret supersymmetric black holes as ensembles of BPS microstates satisfying
a charge constraint that is not respected by individual states. This interpretation provides a microscopic understanding of the hitherto mysterious charge constraints satisfied by all BPS black holes in AdS.
We also develop thermodynamics and a nAttractor mechanism of AdS$_3$ black holes in the nearBPS regime.
}


\arxivnumber{}

\newcommand{\bea}{\begin{eqnarray}}
\newcommand{\eea}{\end{eqnarray}}
\newcommand{\la}{\label}
\newcommand{\be}{\begin{equation}}
\newcommand{\ee}{\end{equation}}

\newcommand{\wt}{\widetilde}

\begin{document}

\maketitle

\section{Introduction}\label{sec:introduction}

The microscopic origin of the Bekenstein-Hawking entropy \cite{Strominger:1996sh}
has been one of the most prominent topics in all of theoretical physics for several decades.
It is largely what triggered the celebrated AdS/CFT correspondence \cite{Maldacena:1997re}
and it continues to serve as an indispensable theoretical laboratory for many aspects of quantum gravity.
However, despite very significant early investigations
\cite{Aharony:2003sx,Kinney:2005ej,Romelsberger:2005eg,Berkooz:2008gc,Chang:2013fba},
only in the last few years was progress made towards understanding the entropy of asymptotically AdS$_{d>3}$ black holes microscopically
\cite{Benini:2015noa,Benini:2015eyy,Benini:2016rke,Hosseini:2017mds,Cabo-Bizet:2018ehj,Choi:2018hmj,Benini:2018ywd,Choi:2019miv,Zaffaroni:2019dhb}.
Moreover, the physical picture behind these recent developments remains blurred by various technicalities even now. 
The purpose of this paper is to exploit well-established insights into black holes in AdS$_3$ to illuminate these conceptual challenges. 

Supersymmetric black holes in AdS$_5\times S^5$, dual to 4D $\mathcal{N}=4$ super-Yang-Mills with $SU(N)$ gauge group, have
entropy that scales as 
$S\sim N^2$ \cite{Gutowski:2004ez,Gutowski:2004yv,Chong:2005da,Chong:2005hr,Kunduri:2006ek,Wu:2011gq,Kim:2006he}. 
The entropy cannot be accounted for by the conventional superconformal index of SYM which has asymptotic 
behavior $O(e^{N^0})$ \cite{Kinney:2005ej,Romelsberger:2005eg}. However, it is now understood that the superconformal index 
grows as $O(e^{N^2})$ \cite{Cabo-Bizet:2018ehj,Choi:2018hmj, Benini:2018ywd} 
(see also \cite{Kim:2019yrz,Cabo-Bizet:2020nkr,Murthy:2020rbd,Agarwal:2020zwm,GonzalezLezcano:2020yeb,Copetti:2020dil,Goldstein:2020yvj})
when studied as a function of complex chemical potentials, rather than real ones.
Moreover, the resulting density of states accounts precisely for the Bekenstein-Hawking entropy of the dual BPS black hole:
\be\label{eqn:Entropy}
  S = 2 \pi \sqrt{Q_1 Q_2 + Q_2 Q_3 + Q_3 Q_1 - \frac{1}{2} N^2(J_1 + J_2)}~,
\ee
where $Q_I$ (with $I=1,2,3$) denote the R-charges (rotations on $S^5$) and $J_i$ (with $i=1,2$) the
angular momenta within AdS$_5$.

The Legendre transform from the canonical (potentials specified) to the microcanonical (charges specified) ensemble can be formulated as an extremization principle for an entropy function \cite{Hosseini:2017mds,Choi:2018hmj} that is necessarily complex. 
Its extremum successfully yields the correct entropy \eqref{eqn:Entropy} but the requirement that it be real 
imposes an \emph{extra constraint} on the black hole charges:
\be\label{eqn:Constraint}
Q_1 Q_2 Q_3 + \frac{1}{2}N^2 J_1 J_2 =
\left( Q_1 + Q_2 + Q_3 + \frac{1}{2}N^2 \right) \left(Q_1 Q_2 + Q_2 Q_3 + Q_3 Q_1 - \frac{1}{2}N^2 (J_1 + J_2) \right)~.
\ee
The physical origin of this constraint is somewhat mysterious, and the way it arises technically is unfamiliar from previous studies of the 
microscopic black hole entropy in other settings. On the other hand, the extra constraint \eqref{eqn:Constraint} is very much 
anticipated from the gravity side where it is satisfied by all BPS black holes in AdS$_5$ 
\cite{Gutowski:2004ez,Gutowski:2004yv,Chong:2005da,Chong:2005hr,Kunduri:2006ek,Wu:2011gq}, 
in addition to the more conventional BPS mass condition 
\be\label{eqn:BPS1}
  M = \sum_{I=1}^3 Q_I + \sum_{i=1}^2 J_i ~. 
\ee
In other words, all black holes that satisfy the mass formula \eqref{eqn:BPS1} also obey the constraint \eqref{eqn:Constraint} \cite{Larsen:2019oll}.

The necessity of angular momentum, the complexification of potentials, and the extra constraint are features of all BPS black holes in AdS$_{d>3}$. 
They may give the impression that BPS black holes in higher dimensional AdS are fundamentally different from their asymptotically flat relatives which are closely related to the BTZ black holes in AdS$_3$. In this article we show that, on the contrary, BPS black holes in AdS$_3$ are very similar to those in AdS$_{d>3}$ and {\it vice versa}. Indeed, most of the material in the paper is not genuinely new, but it has been reworked so
the analysis of AdS$_3$ black holes closely follows contemporary discussions of the higher dimensional case, in an effort to demystify some of the newer developments. 

The AdS$_3$/CFT$_2$ correspondence is simpler, and therefore more transparent, than its
higher dimensional counterparts because:
\begin{itemize}
\item
There are fewer charges.
\item
The charge constraint analogous to \eqref{eqn:Constraint} is linear.
\item
The superconformal algebra in two dimensions factorizes into two independent factors.
\item
There is a powerful tool in CFT$_2$: modular invariance.
\footnote{Interesting modular-like properties of 4D CFT are being studied as well, see \cite{Razamat:2012uv,Gadde:2020bov} for examples.}
\end{itemize}
It is for these reasons that the AdS$_3$ problem has already been ``solved", to a large extent.

We consider general CFT$_2$'s with $(4,4)$ supersymmetry that are not necessarily chiral, we allow distinct levels $k_{R,L}$ in the two sectors. 
In this theory we study the high temperature grand canonical partition function, computed via modular invariance from the vacuum state, and their dual BTZ black holes. From this simple starting point we derive BPS properties of black holes in several ways. 

The most direct approach is to take an appropriate limit of the thermodynamic expressions. This isolates the zero temperature sector. 
However, supersymmetry demands that, in addition, we engage a gauge field for an $SU(2)_R$ symmetry that is interpreted in spacetime 
as rotation on an $S^3$ fibered over AdS$_3$. Thus the BPS limit involves two conditions on the thermodynamic potentials. 

In terms of charges, one of the conditions satisfied by BPS black holes in AdS$_3$ is a linear mass condition that we present as:
\be\label{eqn:BPS1'}
E  - E_{\rm SUSY} = P + J_L~, 
\ee
where $P$ and $J_L=J_1 + J_2$ are conserved charges of the black hole.
The left hand side, including the supersymmetric Casimir energy $E_{\rm SUSY} = - \frac{1}{2}k_L$, corresponds to the black hole mass in the 
higher dimensional examples. We see that the form of the mass formula in AdS$_3$ is completely analogous to \eqref{eqn:BPS1}. 

The second condition satisfied by BPS black holes in AdS$_3$ is a constraint on the black hole charges, namely
\be\label{eqn:BPS2}
  J_L=k_L ~. 
\ee
We interpret this relation as the AdS$_3$ analogue of the constraint \eqref{eqn:Constraint}. Despite its simplicity, it is far from trivial.
The BPS states identified by the superconformal algebra are, in our conventions, the chiral primaries. They all satisfy the mass formula \eqref{eqn:BPS1'} and unitarity further demands that $0\leq J_L \leq 2k_L$ \cite{Eguchi:1987sm,Eguchi:1987wf}. The charge constraint \eqref{eqn:BPS2} is much stronger, it shows that black holes are possible only for a single value of $J_L$. As we explain further below, we interpret this fact as a result of ensemble average. 

Following the cue from recent work on BPS black holes in higher dimensional AdS, we also study the supersymmetric index ${\cal I}$, i.e. the elliptic genus in CFT$_2$.  It is simple to compute via an analytical condition from the partition function and, in the case $k_R=k_L$, we find
\begin{equation}
\label{eqn:lnI}
\ln {\cal I} = k \frac{\tilde{\omega}_1\tilde{\omega}_2}{\tilde{\mu}}~. 
\end{equation}
The variables are potentials that are subject to the constraint
\begin{equation}
\label{eqn:ln2}
\tilde{\mu} - \tilde{\omega}_1 - \tilde{\omega}_2 = 2\pi i ~.
\end{equation}
These formulae give an AdS$_3$ version of the HHZ free energy that plays a central role in discussions of AdS black holes in higher 
dimensions \cite{Hosseini:2017mds}. 
We analyze it by defining the entropy function as a Legendre transform of \eqref{eqn:lnI}, or more precisely its generalization \eqref{eqn:ind} to $k_R\neq k_L$. 
After extremization over all potentials, our entropy function becomes 
\be\la{eqn:Sintro}
S = 2 \pi \sqrt{k_R (P+\frac{1}{2}J_L - \frac{1}{4}k_L) - \frac{1}{4}J_R^2} + \pi i (J_L-k_L)~.
\ee
Upon requiring this to be real, we recover the charge constraint $J_L=k_L$ given in \eqref{eqn:BPS2} and we further find the 
correct BPS entropy
\begin{eqnarray}\la{eqn:BPSSintro}
S_{\rm BPS} & = &  2\pi \sqrt{  k_R (P + \frac{1}{4} k_L) - \frac{1}{4} J^2_R}~.
\end{eqnarray}
The fact that these manipulations are much simpler than their higher dimensional analogues facilitates a critical evaluation of the procedure. 
Alas, we find the reasoning unsatisfying: the imaginary part of \eqref{eqn:Sintro} is immaterial to the reality of physical quantities because 
$J_L - k_L \in \mathbb{Z}$ and so the degeneracy $e^S$ is manifestly real, even before imposing any condition.

In the AdS$_3$ context we can examine {\it why} the manipulations ``work". The real part of the index condition \eqref{eqn:ln2}
indicates that the index does not distinguish the two charges $P$ and $J_L$, it only depends on the combination $P + \frac12 J_L$.
It is extremization over the potentials independently, rather than their combination, that gives the correct charge constraint from a 
principled point of view. That the reality condition gives the same result appears to be an artifact 
of special mathematical properties of the BPS partition function.

Instead, we provide a physical interpretation of the AdS$_3$ charge constraint \eqref{eqn:BPS2}
that is purely microscopic: the \emph{ensemble average}.
While it is not a novel claim that black holes are described by thermal ensembles in the dual field theory,
we show that the very concept of thermal ensemble,
that macroscopic charges are obtained by taking averages over the ensemble,
leads to their constraint.
We expect this observation to be central to understanding more intricate problems in higher 
dimensions, despite its simplicity. 

The rest of this paper is organized as follows.
In section \ref{sec:partfn} we develop the thermodynamics of asymptotically AdS$_3$ BPS black holes with
all chemical potentials treated as real. 
In section \ref{sec:index} we define the supersymmetric index, as opposed to partition function, and potentials become
complex. We formulate an entropy extremization principle and examine why this procedure works. 
We also introduce a nAttractor mechanism for the BTZ black holes, to give a clear spacetime interpretation of the potentials. 
In section \ref{sec:nBPS} we generalize the thermodynamics of the black holes to the nearBPS regime.
Finally, in section \ref{sec:micro}, we discuss how the charge constraint \eqref{eqn:BPS2} arises from an ensemble 
average, by considering the representation theory of $(4,4)$ SCFT$_2$'s.

\section{Partition Function for BTZ Black Holes}\la{sec:partfn}

In this section we study the thermodynamics of BPS black holes in AdS$_3$.
The starting point is the high temperature partition function which we motivate from
both sides of the  AdS$_3$/CFT$_2$ correspondence. We show that the BPS limit
imposes two conditions on the black hole parameters. 

\subsection{Notation}

We consider the standard set-up that describes BPS black holes in 5 asymptotically flat dimensions.
Such black holes lift to the 6D geometry AdS$_3\times S^3$
and are dual to CFT$_2$'s with $(4,4)$ supersymmetry.
The $SU(2)\times SU(2)$ isometry of $S^3$ corresponds to rotation of the original black hole in five dimensions
and is identified with the R-symmetry of the CFT$_2$.

We define the grand canonical partition function as
\begin{equation}
\label{eqn:partdef}
Z = {\rm Tr} ~e^{-\beta(\epsilon - \mu p - \omega_R j_R - \omega_L j_L)}~,
\end{equation}
where the quantum numbers $\epsilon, p, j_R, j_L$ characterize individual states.
The corresponding macroscopic charges, evaluated as averages over many states, are denoted $E, P, J_R, J_L$.
The conjugate potentials of both microscopic and macroscopic quantities are $\beta, \beta\mu, \beta\omega_{R,L}$
with signs specified by the definition (\ref{eqn:partdef}).
Alternatively, the first law of thermodynamics 
$$
T dS = dE - \mu dP  - \omega_R dJ_R - \omega_L dJ_L~, 
$$
summarizes conventions conveniently in a form that is well adapted to black holes. 

In CFT$_2$ the eigenvalues of Virasoro generators are introduced through
$$
L_0 - \frac{k_R}{4} = \frac{\epsilon + p}{2}~, \quad \tilde{L}_0 - \frac{k_L}{4} = \frac{\epsilon - p}{2}~.
$$
The constants $k_{L,R}$ are levels of the $SU(2)$ R-currents. They are related to central charges as $c_{L,R} = \frac{1}{6}k_{L,R}$
by ${\cal N}=4$ supersymmetry.
The unique $SL(2)\times SL(2)$ invariant ground state annihilated by $L_0$, $\tilde{L}_0$ has strictly negative energy $E_{\rm vac}=-\frac{1}{4}(k_R+k_L)$
and corresponds to the AdS$_3$ vacuum. It is separated by a gap from the black holes which have nonnegative 
energy in the CFT$_2$ terminology. The momentum $P$ corresponds to angular momentum of the AdS$_3$ black hole
but for the 5D black hole it is momentum along a compact 6th dimension.

\subsection{The High Temperature Partition Function}

The high temperature partition function plays a central role in our considerations.
In fact, we will regularly refer to it as the ``general" partition function despite the restriction to high temperature, 
in order to stress that it depends on all the continuous variables appearing in the definition \eqref{eqn:partdef}. 
We write it in either of the two forms
\bea
\label{eqn:genlnZ}
\ln Z &=& \frac{k_R}{\beta(1-\mu)} \left( \pi^2 + \beta^2 \omega^2_R\right) + \frac{k_L}{\beta(1+\mu)} \left( \pi^2 + \beta^2 \omega^2_L\right)
\nonumber \\
& = &  \frac{\pi i k_R}{2\tau} \left( 1 - 4 z^2\right) - \frac{\pi i k_L}{2\bar{\tau}}\left( 1 - 4 {\bar z}^2\right) ~.
\eea
The second line is a rewriting of the first that introduces standard CFT$_2$ notation for the fugacities:
\bea
2\pi i \tau &=& - \beta(1 - \mu) ~, \cr
2\pi i \bar{\tau} &=& \beta(1 + \mu)~, \cr
2\pi i z &=& \beta\omega_R ~, \cr
2\pi i \bar{z} &=& - \beta\omega_L~.
\eea
Note that, in either notation, the partition function is a function of four independent real variables.
In contrast, the index corresponds to a boundary condition that sets ${\bar z} = \frac{1}{2}$ and is automatically 
independent of $\bar{\tau}$. Thus the index depends on only two real variables and the dependence on 
the anti-holomorphic ($L$) sector disappears entirely.
We study the index in section \ref{sec:index}.

The simplest derivation of the partition function (\ref{eqn:genlnZ}) applies a modular transformation 
to the ground state contribution. However, the result is very robust and can be reached in many ways. For example, 
a more refined derivation was given in \cite{Kraus:2006nb}, 
from both bulk (AdS$_3$) and boundary (CFT$_2$) points of view. It showed that, when starting from bulk principles, 
all (local) higher derivative corrections are incorporated.  

From the general partition function \eqref{eqn:genlnZ},
thermodynamic properties such as macroscopic variables of the ensemble are readily obtained.
Differentiation of the partition function (\ref{eqn:genlnZ}) by $\beta$ gives 
\bea\la{eqn:E'gen}
E -\mu P - \omega_R J_R - \omega_L J_L & = & - \frac{\partial\ln Z}{\partial\beta} \nonumber \\
& = & 
\frac{k_R}{\beta^2(1-\mu)} \left( \pi^2 - \beta^2 \omega^2_R\right) +
\frac{k_L}{\beta^2(1+\mu)} \left( \pi^2 - \beta^2 \omega^2_L\right)~,
\eea
and we similarly find the conserved charges
\begin{eqnarray}
\label{eqn:Pgen}
P &=& \frac{1}{\beta}\frac{\partial\ln Z}{\partial\mu} = 
\frac{k_R}{\beta^2(1-\mu)^2} \left( \pi^2 + \beta^2 \omega^2_R\right) - \frac{k_L}{\beta^2(1+\mu)^2} \left( \pi^2 + \beta^2 \omega^2_L\right)
~,
\\
\label{eqn:JRLgen}
J_{L,R} & = & \frac{1}{\beta}\frac{\partial\ln Z}{\partial\omega_{L,R}}
= \frac{2k_{L,R}}{1\pm\mu}  \omega_{L,R}~.
\end{eqnarray}
A combination of these expressions gives the energy
\begin{equation}\label{eqn:Egen}
E = \frac{k_R}{\beta^2(1-\mu)^2} \left( \pi^2 + \beta^2 \omega^2_R\right)
+    \frac{k_L}{\beta^2(1+\mu)^2} \left( \pi^2 + \beta^2 \omega^2_L\right)~,
\end{equation}
and the macroscopic entropy 
\begin{eqnarray}\label{eqn:Sgen}
S &=& \beta\left( E - \mu P - \omega_R J_R - \omega_L J_L\right) + \ln Z \nonumber \\
&=& \frac{2k_R \pi^2}{\beta(1-\mu)}  + \frac{2k_L \pi^2}{\beta(1+\mu)}
\nonumber \\
& = &  2\pi \sqrt{ \frac{1}{2} k_R(E+P) - \frac{1}{4} J^2_R} + 2\pi \sqrt{ \frac{1}{2} k_L (E- P) - \frac{1}{4} J^2_L}~.
\end{eqnarray}
Equations (\ref{eqn:Pgen}-\ref{eqn:Sgen}) are starting points for various limits we study in the rest of this section.

\subsection{Supersymmetry Gives Two Conditions on Parameters}
\label{subsec:BPSdis}

Up to this point we did not impose any conditions on the black hole parameters.
We now impose supersymmetry and show that the resulting BPS black holes satisfy {\it two} conditions. 

In the 2D superconformal theory with $(4,4)$ supersymmetry, there are four $\frac14$-BPS sectors.
Each sector preserves two real supersymmetries that are either holomorphic ($R$) or anti-holomorphic ($L$),
and that either raise or lower the R-charge.
We focus without loss of generality throughout the article to the $\frac14$-BPS sector which preserves
supersymmetries that are anti-holomorphic ($L$) and raise the R-charge.
Then the unitarity bound from the anticommutator of the supercharges on individual CFT states in the
NS sector is:
\be\la{eqn:unitboundmicro}
\epsilon - p + \frac{1}{2}k_L\geq j_L  ~,
\ee
from which a bound for black hole energy and charges follows:
\be\la{eqn:unitboundmacro}
 E - P + \frac{1}{2}k_L\geq J_L  ~.
\ee
Microscopic states whose quantum numbers saturate the inequality \eqref{eqn:unitboundmicro} are called chiral primaries.
Unitarity further requires that chiral primaries have $0\leq j_L \leq 2k_L$ \cite{Eguchi:1987sm,Eguchi:1987wf}.

Saturation of the inequality \eqref{eqn:unitboundmacro} is a necessary condition for a supersymmetric black hole 
but it is not sufficient. Indeed, the black hole entropy formula \eqref{eqn:Sgen} does not make sense unless
\cite{Cvetic:1998xh}: 
\be\la{eqn:regbound}
\frac{1}{2} ( E - P )  \geq \frac{1}{4k_L}J^2_L  ~.
\ee
A hypothetical black hole solution that violates this inequality would have event horizon with imaginary area. Such geometries 
are not regular so black holes with these quantum numbers simply do not exist. This regularity condition is variously referred to as
the cosmic censorship bound or the condition for absence of closed time-like curves. 

The BPS condition demands that the inequality \eqref{eqn:unitboundmacro} be saturated but then compatibility with regularity 
\eqref{eqn:regbound} gives
\be\la{eqn:JL=kL}
J_L=k_L~.
\ee
This is the charge constraint on BPS black holes in AdS$_3$ advertised in the introduction \eqref{eqn:BPS2}.
Thus BPS black holes have the same quantum numbers as the particular
chiral primaries situated in the middle of the interval  $0\leq j_L \leq 2k_L$ allowed by unitarity. 

\subsection{Extremality vs. Supersymmetry}\la{subsec:codim2}

In the previous subsection we established that BPS black holes in AdS$_3$ are co-dimension 2 in parameter space:
saturation of {\it two} inequalities (\ref{eqn:unitboundmacro}-\ref{eqn:regbound})
introduces {\it two} relations between the four parameters $E$, $P$, and $J_{R,L}$.
In this and the next subsection we elaborate on this property from a thermodynamic point of view.

In discussions of black holes two notions of ``ground state" appear:
\begin{itemize}
\item
\emph{Extremality}: the temperature $T=0$. 
\item
\emph{Supersymmetry}: the BPS inequality for the energy is saturated. 
\end{itemize}
These conditions are similar in that both determine the black hole energy in terms of its charges. However, they are not at all equivalent. 
On the contrary, it may be useful to interpret them as two complementary requirements that each imposes one relation between the black hole 
parameters. The supersymmetric black holes are co-dimension $2$ in parameter space because of these two conditions.
\footnote{In this paper we just consider conditions on continuous black hole parameters. There are also important discrete distinctions that must be made, such as the ones defining the nonBPS branch \cite{Gimon:2007mh}.}

The two concepts of ground state can be applied in either order. In the previous subsection our starting point was the supersymmetry algebra: 
\begin{enumerate}
\item
Supersymmetry gives the BPS condition $E = P + J_L - \frac12 k_L$ that determines the energy $E$ in terms of conserved charges. In CFT$_2$ terminology the eigenvalue of $L_0$ is $\frac{1}{2}J_L$. 
\item
Among configurations with charges that satisfy the BPS formula for the energy, a regular black hole exists only if, in addition, the extremality condition 
$$
T^{-1} = \beta = \left( \frac{\partial S}{\partial E}\right)_{P, J_{L,R}} 
= \frac{k_L \pi}{2\sqrt{\frac{k_L}{2} ( E - P )  - \frac{1}{4}J^2_L}}  + \frac{k_R\pi}{2\sqrt{\frac{k_R}{2} ( E + P )  - \frac{1}{4}J^2_R}} \to\infty~,
$$
%
is met. This is only possible when the charges are further restricted to $J_L = k_L$. 
\end{enumerate} 
From this point of view the second condition on charges is ``additional" and perhaps surprising. 
However, thermodynamic reasoning suggests that we impose extremality first: 
\begin{enumerate}
\item
The lowest possible energy allowing a regular black hole geometry for given conserved charges $(P, J_{L,R})$ is the extremal energy
$E_{\rm ext}$.
\item
Considering only extremal black holes, we further require that the geometry permits supersymmetry: a spacetime Killing spinor must exist. This imposes an independent constraint on the charges.   
\end{enumerate}
From the thermodynamic point of view it is supersymmetry that imposes an 
additional condition on the charges that may appear surprising. 
In the next subsection we will implement the BPS limit with extremality imposed first. 
In particular, we will derive the two inequalities (\ref{eqn:unitboundmacro}-\ref{eqn:regbound})
defining the BPS limit from the general partition function \eqref{eqn:genlnZ}.

\subsection{BPS as a Thermodynamic Limit}\label{sec:BPSthermo}

Recall the formulae (\ref{eqn:Pgen}-\ref{eqn:Egen}) that relate the quantum numbers to potentials,
reproduced here for convenience:
\begin{subequations}\la{eqn:chargerep}
\bea
\la{eqn:Egen'}
E &=& \frac{k_R}{\beta^2(1-\mu)^2} \left( \pi^2 + \beta^2 \omega^2_R\right)
+    \frac{k_L}{\beta^2(1+\mu)^2} \left( \pi^2 + \beta^2 \omega^2_L\right)~,
\\
P &=& \frac{k_R}{\beta^2(1-\mu)^2} \left( \pi^2 + \beta^2 \omega^2_R\right) - \frac{k_L}{\beta^2(1+\mu)^2} \left( \pi^2 + \beta^2 \omega^2_L\right)
~,
\\
\la{eqn:JRLgen'}
J_{L,R} & = & \
 \frac{2k_{L,R}}{1\pm\mu}  \omega_{L,R}~.
\eea
\end{subequations}
In the canonical ensemble the extremal limit amounts to vanishing temperature $\beta\to\infty$. However, we must 
be careful with what remains finite in this limit.

Consider a pair of particular combinations of these charges:
\begin{subequations}\label{eqn:extbound0}
\bea
\label{eqn:extbound1}
E + P - \frac{J_R^2}{2k_R} &=& \frac{2k_R \pi^2 }{\beta^2 (1-\mu)^2} \geq 0 
~,
\\
\label{eqn:extbound2}
E - P - \frac{J_L^2}{2k_L} &=& \frac{2k_L \pi^2}{\beta^2 (1+\mu)^2} \geq 0
~.
\eea
\end{subequations}
If one na\"ively takes $\beta\to\infty$ with the chemical potential $\mu$ finite and generic,
both of these inequalities will be saturated. However, when the expressions on the left hand sides of
both equations in \eqref{eqn:extbound0} vanish, the black hole entropy \eqref{eqn:Sgen} will be zero as well.
Therefore, the limit taken this way yields an extremal ``black hole" 
with an event horizon that has vanishing area. Such a geometry is singular, it is not a black hole solution. 
 
In order to circumvent this obstacle, we need to saturate only one of the inequalities \eqref{eqn:extbound0}.
We pick the latter without loss of generality, because this choice is analogous to the one leading to \eqref{eqn:regbound}.
Accordingly, we take $\beta\to\infty$ while rescaling $\mu$ so that $\tilde\mu \equiv \beta (\mu-1)$ remains finite.
Note that $\tilde{\mu}\leq 0$ because $\mu\leq 1$.
It further follows from \eqref{eqn:JRLgen'} that, in order to describe black holes with generic values of $J_R$, we 
must further take $\omega_R\to 0$  so that $\tilde\omega_R \equiv \beta \omega_R$ is also kept finite.
In contrast, $\omega_L$ does not require any rescaling, it can be kept finite by itself.

In summary, the extremal limit of a general AdS$_3$ black hole is: 
\be\la{eqn:extlimit}
\text{Extremal limit: } \quad
\begin{cases}
\beta \to \infty ~,& \\
\mu \to 1 & ~ \text{with}~  \tilde\mu \equiv \beta (\mu-1) ~\text{finite,} \\
\omega_R \to 0 & ~ \text{with}~ \tilde\omega_R \equiv \beta \omega_R ~\text{finite,} \\
\omega_L  ~\text{finite.} &
\end{cases}
\ee
This limit was designed so that (\ref{eqn:chargerep}) gives expressions that are finite:
\begin{subequations}\la{eqn:extcharge}
\bea
E &=& 
\frac{k_R}{\tilde{\mu}^2} \left( \pi^2 + \tilde{\omega}^2_R\right) + \frac{k_L}{4} \omega^2_L~,
\\
P &=& 
\frac{k_R}{\tilde{\mu}^2} \left( \pi^2 + \tilde{\omega}^2_R\right) - \frac{k_L}{4} \omega^2_L
~,
\\
J_{R} & = &  - \frac{2k_{R}}{\tilde{\mu}}  \tilde{\omega}_{R}
~,
\\
J_{L} & = &  k_{L}\omega_{L}
~.
\eea
\end{subequations}
The explicit sign in the formula for $J_R$ compensates $\tilde{\mu}<0$ so that the angular momentum $J_R$ 
has the same sign as the rescaled angular velocity $\tilde{\omega}_{R}$, 
as expected. These formulae for the conserved charges give the energy as a function of the charges
\begin{equation}
\label{eqn:Eext}
E_{\rm ext} =  P + \frac{1}{2k_L} J^2_L~.
\end{equation}
This is the ground state energy for these conserved charges. It saturates \eqref{eqn:regbound} and 
is identified with the extremal black hole mass.
The extremal entropy becomes
\begin{eqnarray}\la{eqn:Sext}
S_{\rm ext} &=& - \frac{2k_R \pi^2}{\tilde{\mu}}  
 = 2\pi \sqrt{ \frac{1}{2} k_R(E_{\rm ext}+P) - \frac{1}{4} J^2_R}  \nonumber \\
& = &  2\pi \sqrt{  k_R P + \frac{k_R}{4k_L} J^2_L - \frac{1}{4} J^2_R}~.
\end{eqnarray}
The last equation eliminated the energy using the extremality condition (\ref{eqn:Eext}).

As we have stressed, the extremal black holes are not necessarily supersymmetric. 
As the second and last step of implementing the BPS limit, we now examine supersymmetry.
Recall from \eqref{eqn:unitboundmacro} that charges of supersymmetric black holes must saturate the inequality
$$
 E - P - J_L + \frac{1}{2}k_L\geq 0  ~.
$$
The left hand side can be recast as a sum of two squares  
\be\la{eqn:susybound}
E - P - J_L + \frac{1}{2}k_L = \frac{2k_L \pi^2}{\beta^2 (1+\mu)^2} + \frac{k_L}{2} (1-\frac{2\omega_L}{1+\mu})^2~,
\ee
using (\ref{eqn:chargerep}).
The first square is precisely \eqref{eqn:extbound2} so it vanishes in the extremal limit.
In order to saturate the BPS bound \eqref{eqn:unitboundmacro} the second square must vanish as well so we
demand that the potentials satisfy
\be\la{eqn:BPSpotentials}
\varphi \equiv  1+\mu - 2\omega_L = 0 ~,
\ee
in addition to conditions for extremality. We defined the parameter $\varphi$ for future use. 
Since $\mu = 1$ at extremality we must have $\omega_L = 1$ in the BPS limit. 
However, just as the extremal limit is taken with $\tilde\mu \equiv \beta (\mu -1)$ kept finite there is no obstacle to 
taking the BPS limit $\omega_L \to 1$ so $\tilde{\omega}_L \equiv \beta(\omega_L-1)$ remains finite.
The value of $\tilde\omega_L$ is, like $\tilde\mu$ and $\tilde\omega_R$, not constrained.

To summarize, the BPS AdS$_3$ black holes are limits of generic AdS$_3$ black holes as
\begin{equation}
\label{eqn:BPSlimitT}
T = \beta^{-1} \to 0~,
\end{equation}
while the potentials
\begin{equation}
\label{eqn:BPSlimit}
\tilde\mu = \beta (\mu -1)~, \quad \tilde{\omega}_R = \beta \omega_R~, \quad \tilde{\omega}_L = \beta(\omega_L-1)~,
\end{equation}
are kept finite. 
In this limit two inequalities \eqref{eqn:unitboundmacro} and \eqref{eqn:regbound} are saturated.

The BPS limit of the extremal expressions \eqref{eqn:extcharge} gives
\begin{subequations}\la{eqn:BPScharge}
\bea
E &=& 
\frac{k_R}{\tilde{\mu}^2} \left( \pi^2 + \tilde{\omega}^2_R\right) + \frac{k_L}{4}~,
\\
P &=& 
\frac{k_R}{\tilde{\mu}^2} \left( \pi^2 + \tilde{\omega}^2_R\right) - \frac{k_L}{4}~,
\\
J_{R} & = &  - \frac{2k_{R}}{\tilde{\mu}}  \tilde{\omega}_{R} ~,
\eea
\end{subequations}
and notably,
\bea\la{eqn:BPSJL}
J_{L} & = &  k_{L} ~.
\eea
The extremal black hole entropy \eqref{eqn:Sext} also simplifies further in the BPS limit
\begin{eqnarray}\la{eqn:BPSS}
S_{\rm BPS} & = &  2\pi \sqrt{  k_R (P + \frac{1}{4} k_L) - \frac{1}{4} J^2_R}~.
\end{eqnarray}

The four macroscopic quantities $E, P, J_{L,R}$ are parametrized by only two potentials $\tilde\mu$ and $\tilde\omega_R$, they are independent of 
the third potential $\tilde\omega_L$. This confirms the expectation that the parameters of a BPS black hole form a 
co-dimension 2 surface in the space of all possible charges. On the other hand, there really are three independent rescaled potentials 
$\tilde\mu, \tilde\omega_{L,R}$. This is possible because $\tilde\omega_L$ parametrizes a flat direction along which the BPS black hole does not change.

\subsection{The BPS Limit and the Partition Function}\label{sec:BPSpf}
We now implement the BPS limit discussed in the previous subsection on 
the partition function rather than the macroscopic variables. 

As before, we first take the extremal (zero temperature) limit $\beta\to\infty$ in the manner specified in \eqref{eqn:extlimit}. 
The trace \eqref{eqn:partdef} that defines the partition function becomes
\begin{equation}
\label{eqn:Zext}
Z = {\rm Tr} ~e^{-\beta(\epsilon -p) + \tilde{\mu} p + \tilde{\omega}_R j_R + \beta\omega_L j_L}~.
\end{equation}
This expression is schematic because $\beta$ appears explicitly even though we take $\beta\to\infty$. However, 
it captures an important qualitative feature of the physics.
Disregarding temporarily the term $\beta\omega_L j_L$ (which will be addressed below), as $\beta\to\infty$ the first term in 
the exponent assures that only states with $\epsilon=p$ contribute insofar as such states exist and they are separated from the states with $\epsilon>p$ by a gap. The states singled out this way will be the BPS states, except for the proviso that we have yet to account for the term
$\beta\omega_L j_L$. 

To do so we proceed and implement the second part of the BPS prescription (\ref{eqn:BPSlimitT}-\ref{eqn:BPSlimit}) which specifies the
BPS energy. It is taken into account by rewriting the extremal partition function (\ref{eqn:Zext}) as
\begin{eqnarray}
\label{eqn:ZBPS}
Z &=& e^{\frac{1}{2}\beta k_L} {\rm Tr} ~e^{-\beta(\epsilon -p - j_L + \frac{1}{2}k_L)    + \tilde{\mu} p + \tilde{\omega}_R j_R  +\tilde{\omega}_L  j_L}\nonumber \\
&=& e^{\frac{1}{2}\beta k_L}  {\rm Tr} ~e^{-2\beta(\tilde{L}_0 -\frac{1}{2} j_L)} e^{\tilde{\mu} p + \tilde{\omega}_R j_R  +\tilde{\omega}_L  j_L}~.
\end{eqnarray}
In the second expression we introduced $\tilde{L}_0 - \frac{k_L}{4} = \frac{1}{2} ( \epsilon -p)$ and reorganized in order
to isolate the term $\beta(\tilde{L}_0 -\frac{1}{2} j_L)$ in the exponent which, because the limit $\beta\to\infty$ is implied, singles out the chiral primary states 
annihilated by $\tilde{L}_0 -\frac{1}{2} j_L$. 
We assume that such states are separated by a gap from the states where $\tilde{L}_0 -\frac{1}{2} j_L$  is positive and
unitarity ensures that this operator cannot be negative. Thus the partition function 
receives contributions only from the chiral primaries, precisely the states that preserve supersymmetry.

The overall factor $e^{\frac{1}{2}\beta k_L}$ in \eqref{eqn:ZBPS} diverges as $\beta\to\infty$ but, because no other potential enters,
it does not depend on the state. This term incorporates the supersymmetric Casimir energy \cite{Assel:2015nca}
\begin{equation}
\label{eqn:SUSYCast}
E_{\rm SUSY} = - \frac{1}{2}k_L~,
\end{equation}
that is common to all states. Note that it is not the regular Casimir energy $E_C = - \frac{1}{4} ( k_L + k_R)$ that enters here and the two notions of Casimir energy agree only when the levels $k_L=k_R$. The Casimir energy appears explicitly because we study
the partition function {\it  defined as a path integral} rather than as a trace over a Hilbert space normalized such that the vacuum contributes unity.

It is the convention in CFT$_2$ that the Virasoro generators $L_0$, $\tilde{L}_0$ annihilate the $SL(2)^2$ invariant (NS-NS)-vacuum which, therefore, is assigned a negative Casimir energy $E_C = - \frac{1}{24}(c_R + c_L) = - \frac{1}{4}(k_R + k_L)$. 
This usage has been adopted in discussions of AdS$_3$/CFT$_2$ correspondence. The 
supersymmetric Casimir energy \eqref{eqn:SUSYCast} is a variant that is better protected by supersymmetry, but it follows the same conventions. In contrast, in the context of black holes in higher dimensional AdS spaces, it is customary to assign mass $M=0$ to the AdS vacuum. Adaptation of our AdS$_3$ treatment to this practice amounts to defining the BPS black hole mass as
\begin{eqnarray}\la{eqn:BPSE}
M = E  - E_{\rm SUSY}  &=&  P   + J_L ~.
\end{eqnarray}
This simple linear formula, with numerical value ``1" in front of each quantum number $P$ and $J_L$, is the AdS$_3$ version of the standard supersymmetric mass formulae for supersymmetric black holes in AdS$_{4, 5, 6, 7}$. 

Taking the extremal limit \eqref{eqn:extlimit} explicitly on the general partition function (\ref{eqn:genlnZ}) we find
\begin{equation}
\label{eqn:extlnZ}
\ln Z_{\rm ext} = - \frac{k_R}{\tilde{\mu}} \left( \pi^2 + \tilde{\omega}^2_R\right) + \frac{1}{2} k_L  (\beta  - \frac{1}{2}\tilde{\mu}) \omega^2_L~.
\end{equation}
We retained the divergent linear-in-$\beta$ term which encodes the supersymmetric Casimir energy but does not contribute to the entropy. Other terms were computed by expanding for small temperature and
retaining the terms that are finite in the extremal limit.
The extremal partition function (\ref{eqn:extlnZ}) simplifies further in the BPS limit 
\begin{equation}
\label{eqn:lnZBPS}
\ln Z_{\rm BPS} = \frac{1}{2} k_L  \beta - \frac{k_R}{\tilde{\mu}} \left( \pi^2 + \tilde{\omega}^2_R\right) +
k_L  \left( \tilde{\omega}_L - \frac{1}{4} \tilde{\mu} \right) ~.
\end{equation}
This BPS partition function 
reproduces the formulae for BPS limits of macroscopic charges (\ref{eqn:BPScharge}-\ref{eqn:BPSJL}). For example,
the potential $\tilde{\omega}_L$ now appears entirely as a linear term that gives the correct value
\be\la{eqn:BPSJL2}
J_L = \frac{\partial}{\partial \tilde\omega_L} \ln Z_{\rm BPS} = k_L~.
\ee

\section{The Supersymmetric Index and Entropy Extremization}\la{sec:index}

In the previous section we discussed black hole thermodynamics with the partition function
as starting point, as in conventional thermodynamics. However, recent progress on BPS black holes in AdS with 
dimensions larger than three is based on the superconformal index. Therefore, in this section, we study the thermodynamics of BTZ black holes 
on the basis of the supersymmetric index. 
In particular, we develop an entropy extremization prescription for BTZ black holes that mimics its analogues in the literature 
on higher dimensional cases \cite{Hosseini:2017mds}. 

\subsection{The Partition Function and the Index}\label{ssec:pfind}

The grand canonical partition function was defined in (\ref{eqn:partdef}), as a trace over all states: 
\begin{eqnarray}
\label{eqn:partdefre}
Z &=& {\rm Tr} ~e^{-\beta(\epsilon - \mu p - \omega_R j_R - \omega_L j_L)}~.
\end{eqnarray}
In subsection \ref{sec:BPSpf}  we isolated the BPS states by taking $\beta \to \infty$ with certain rescaled potentials (identified by their tilde) 
kept finite. This gave the BPS partition function (\ref{eqn:ZBPS}): 
\begin{eqnarray}
\label{eqn:ZBPSrep}
Z_{\rm BPS} &=& \left. e^{\frac{1}{2}\beta k_L}  {\rm Tr} ~e^{-2\beta(\tilde{L}_0 -\frac{1}{2} j_L)} e^{\tilde{\mu} p + \tilde{\omega}_R j_R  +\tilde{\omega}_L  j_L} \right|_{\beta \to \infty} \nonumber \\
&=& \left. e^{\frac{1}{2}\beta k_L} \right|_{\beta \to \infty}
{\rm Tr}_{\rm BPS} ~ e^{\tilde{\mu} p + \tilde{\omega}_R j_R +\tilde{\omega}_L  j_L} ~.
\end{eqnarray}
The limit $\beta\to\infty$ ensures that only the chiral primaries contribute to the trace since
the operator $\tilde{L}_0 -\frac{1}{2} j_L$ vanishes exactly on those and is positive on others.
Equivalently, the trace is taken only over the chiral primaries (BPS states) in the second line.

In this section we study the supersymmetric index, also known as the elliptic genus in CFT$_2$, rather than the partition function. As usual, 
the index is the general partition function \eqref{eqn:partdefre}, except for insertion into the trace of a sign $(-1)^F$ that depends on the fermion number $F$. The goal is that when the supercharge ${\cal Q}$ that defines the BPS sector does not annihilate some state $|\psi\rangle$, 
it creates a nontrivial partner ${\cal Q}|\psi\rangle$ that cancels the original state $|\psi\rangle$ in the trace, because the two members of the pair 
are counted with opposite signs $(-1)^F$. The general partition function \eqref{eqn:partdefre} with $(-1)^F$ inserted should therefore receive contributions 
only from states that are annihilated by ${\cal Q}$ and so reduce to the 
BPS partition function \eqref{eqn:ZBPSrep}, also with $(-1)^F$ inserted. 

However, for the two members of each pair to cancel properly, they must have the same fugacities,
their weight depending on the potentials with tilde must be the same.
This can be arranged by considering only fugacities that satisfy the constraint
\be\la{eqn:EGcond}
\beta(1+\mu - 2\omega_L ) = \tilde\mu - 2\tilde\omega_L = 0~,
\ee
which commutes with the supercharge ${\cal Q}$ in the anti-holomorphic ($L$) sector. More concisely, the insertion of $(-1)^F$ and the requirement $\tilde\mu - 2\tilde\omega_L=0$ can be elegantly combined as the complex constraint
\be\la{eqn:indcond}
\tilde\mu - 2\tilde\omega_L = 2 \pi i ~,
\ee
on the potentials.
With this constraint the general partition function \eqref{eqn:partdefre} automatically reduces to 
the BPS partition function \eqref{eqn:ZBPSrep}. In particular, the dependence on $\beta$ disappears, 
except for the factor $e^{\frac12 \beta k_L}$ that accounts for the supersymmetric Casimir energy. 
It is conventional to omit this overall factor from definitions of supersymmetric indices, or of elliptic genus.

To summarize,
\begin{eqnarray}
\label{eqn:defI}
{\cal I} &\equiv& \left. e^{\beta E_{\rm SUSY}} Z
\right|_{\tilde\omega_L = \frac{\tilde\mu}{2} - i \pi} \nonumber \\
&=& \left. {\rm Tr}_{\rm BPS}~e^{\tilde{\mu}p +\tilde{\omega}_R j_R +\tilde{\omega}_L j_L} 
\right|_{\tilde\omega_L = \frac{\tilde\mu}{2} - i \pi} \nonumber \\
&=& \left. e^{\beta E_{\rm SUSY}} Z_{\rm BPS}
\right|_{\tilde\omega_L = \frac{\tilde\mu}{2} - i \pi}~,
\end{eqnarray}
where $E_{\rm SUSY}= -\frac12 k_L$ was given in \eqref{eqn:SUSYCast}.
Going from the first to the third line is non-trivial,
it is valid because the aforementioned cancellations within pairs allow one to restrict the trace to BPS states.
In other words, the index is independent of $\beta$, as expressed by the second line of \eqref{eqn:defI}, 
so $\beta \to \infty$ is not needed in its definition.

The BPS partition function $Z_{\rm BPS}$ depends on \emph{three} independent potentials:
$\tilde\mu$ and $\tilde\omega_{L,R}$, apart from the formal
$\left. e^{-\beta E_{\rm SUSY}} \right|_{\beta \to \infty}$ factor.
Since the dependence on $\tilde\omega_L$ can be eliminated by the complex constraint (\ref{eqn:indcond}),
the index depends on only \emph{two} independent parameters which we take as $\tilde\mu$ and $\tilde\omega_R$.

We can compute the index for supersymmetric black holes in AdS$_3$ explicitly by starting from the general partition function (\ref{eqn:genlnZ}), 
introducing tilde potentials through \eqref{eqn:BPSlimit}, and then imposing the index constraint (\ref{eqn:indcond}):
\bea
\label{eqn:ind}
\ln {\cal I} &=& - \frac{k_L}{2}\beta  +   \frac{k_R}{\beta(1-\mu)} \left( \pi^2 + \beta^2 \omega^2_R\right) + \frac{k_L}{\beta(1+\mu)} \left( \pi^2 + \beta^2 \omega^2_L\right) \nonumber \\
&=& - \frac{k_L}{2}\beta -\frac{k_R}{\tilde\mu} \left( \pi^2 + \tilde\omega^2_R\right) + \frac{k_L}{\tilde\mu + 2\beta} \left(\pi^2 + (\tilde\omega_L +\beta)^2\right) \nonumber \\
&=& -\frac{k_R}{\tilde\mu} \left( \pi^2 + \tilde\omega^2_R\right) + \frac{k_L}{4} (\tilde\mu -4 \pi i ) 
\nonumber \\
&=&-\frac{k_R}{\tilde\mu} \left( \pi^2 + \tilde\omega^2_R\right) + \frac{k_L}{\tilde\mu} (\pi^2+\tilde\omega_L ^2 )~.
\eea
We present the manipulations in detail to highlight that they are exact and
that the dependence on $\beta$ disappears without any limit taken, as anticipated.
The final expression with the constraint (\ref{eqn:indcond}) implied
agrees with the BPS partition function (\ref{eqn:lnZBPS}), again as anticipated.
A simpler but less illuminating route to the formula for the index given in the last line of \eqref{eqn:ind} is to evaluate 
the partition function and take the {\it high} temperature limit $\beta\to 0$ with the tilde variables kept fixed. In other words, the
last line of \eqref{eqn:ind} follows from the second line by taking $\beta=0$. 

The computation illustrates how the index \eqref{eqn:defI} and the BPS partition function (\ref{eqn:lnZBPS}) are closely related, 
yet they are different in significant ways such that they complement one another:
\begin{itemize}
\item
The BPS partition function restricts the trace to the chiral primary states by an explicit limit $\beta \to \infty$.
In contrast, the index is independent of $\beta$, the limit $\beta \to \infty$ is possible but not mandatory.
This is one aspect of the index being protected under continuous deformations of the theory, while the BPS partition function is not.
\item
The supersymmetric index is defined not only by an insertion of $(-)^F$, its fugacities must be constrained by (\ref{eqn:indcond}) or else it is not protected under continuous deformations. In contrast, the BPS partition function 
keeps all three potentials $\tilde\mu$ and $\tilde\omega_{R,L}$ independent. It is possible to focus on variables that satisfy the constraint, but
the general case incorporates more information about the theory. 
\item
The supersymmetric index is defined with the supersymmetric Casimir energy stripped off, while the partition function retains it. 
\end{itemize}
These distinctions between the supersymmetric index and the BPS partition function are central to this paper. 

In the non-chiral case $k_L=k_R=k$ we can recast our result for the index \eqref{eqn:ind} as
\be\la{eqn:lnZindex2}
 \ln {\cal I} = k \frac{\tilde{\omega}_1\tilde{\omega}_2}{\tilde{\mu}}~, 
\ee
by choosing the basis $\tilde{\omega}_{L,R} = \frac{1}{2} ( \tilde{\omega}_1 \pm  \tilde{\omega}_2)$ 
for the potentials. 
This result is reminiscent of the HHZ free energy that plays a central role
in discussions of black hole entropy in higher dimensional AdS spaces.
For example, in AdS$_5$/CFT$_4$ \cite{Hosseini:2017mds},
\be\la{eqn:lnZ5}
\ln Z_5  = \frac{1}{2} N^2  \frac{\tilde{\Delta}_1\tilde{\Delta}_2\tilde{\Delta}_3}{\tilde{\omega}_a\tilde{\omega}_b}~.
\ee
The three potentials $\tilde\Delta_I$ ($I=1,2,3$) for R-charges in the higher dimensional setting (rotation on $S^5$)
are analogous to $\tilde{\omega}_{1,2}$ for R-charges in CFT$_2$ (rotation on $S^3$).
The rotational velocities $\tilde{\omega}_{a,b}$ (not to be confused with $\tilde{\omega}_{1,2}$ in \eqref{eqn:lnZindex2})
in AdS$_5$ correspond to the potential for angular momentum $\tilde{\mu}$ in AdS$_3$.
The overall coefficient $k$ is two times the Casimir energy in AdS$_3$ while $\frac{1}{2} N^2$ is 
two times the Casimir energy in AdS$_5$.

We interpret our result for the supersymmetric index \eqref{eqn:ind} as the HHZ free energy
in AdS$_3$.
It is more general than the version \eqref{eqn:lnZindex2} that is more directly analogous to
the HHZ formulae in higher dimensions, because it includes the non-chiral case $k_R\neq k_L$.
In each dimension, the index nature of the HHZ free energy requires imposing a
linear constraint between the complexified potentials: 
the 3D free energy \eqref{eqn:ind} satisfies \eqref{eqn:indcond} and the constraint
$$
\tilde{\Delta}_1+\tilde{\Delta}_2+\tilde{\Delta}_3-\tilde{\omega}_a-\tilde{\omega}_b=2\pi i~,
$$
is imposed on the 5D free energy \eqref{eqn:lnZ5}. In the AdS$_3$ example we can make completely explicit
the distinction between the HHZ free energy \eqref{eqn:ind} and the BPS partition function \eqref{eqn:lnZBPS} that depends on unconstrained potentials.
This comparison also highlights the role of the supersymmetric Casimir energy.

\subsection{Entropy Extremization}\label{sec:entext}

Whereas we have derived the supersymmetric index (\ref{eqn:ind}) for AdS$_3$ black holes
by imposing a complex condition (\ref{eqn:indcond}) on the more general BPS partition function,
in higher dimensional AdS spaces it is only the index that can be reliably computed.
In that context a procedure to extract the entropy and the charge constraint
of supersymmetric black holes directly from the index has been developed \cite{Hosseini:2017mds}.
In this subsection we apply this procedure to the AdS$_3$ case and show that it
reproduces the results derived from the BPS partition function in section \ref{sec:partfn}. 

The claim that is now standard in higher dimensional AdS spaces is that we can process the index as if it was an ordinary free 
energy. It is with this procedure in mind that we have referred to the (logarithm of the) index as the HHZ free energy.
According to this prescription, the black hole entropy is given by the Legendre transform of the index (\ref{eqn:ind}), 
subject to the complex constraint (\ref{eqn:indcond}).
Following \cite{Hosseini:2017mds}, it can be computed efficiently by extremizing the entropy function
\be\la{eqn:Sfunction}
S [ \tilde\mu, \tilde\omega_{R} , \tilde\omega_{L} ] = \frac{k_L \left( \tilde{\omega}_L^2 + \pi^2\right) - k_R \left( \tilde{\omega}_R^2 + \pi^2\right)}{\tilde{\mu}}
 -  \tilde{\omega}_L J_L  -  \tilde{\omega}_R J_R - \tilde{\mu} P - \Lambda (\tilde{\mu} - 2\tilde{\omega}_L - 2\pi i)~,
\ee
with respect to the potentials $\tilde\mu$, $\tilde\omega_{R,L}$ and the Lagrange multiplier $\Lambda$ that 
enforces the condition \eqref{eqn:indcond}.

The entropy function is homogeneous of degree one in the potentials $\tilde\mu$, $\tilde\omega_{R,L}$, except 
for $2\pi i \Lambda$ which is constant, and for the terms proportional to $\pi^2$ which are homogeneous of degree minus one. 
Keeping track of the inhomogeneous terms, the extremization conditions give
$$
0 = \left( \tilde{\omega}_L\partial_{\tilde{\omega}_L}  + \tilde{\omega}_R\partial_{\tilde{\omega}_R}  + \tilde{\mu}\partial_{\tilde{\mu}}\right)S  =S - 2\pi i \Lambda + \frac{2\pi^2 (k_R-k_L)}{\tilde{\mu}}~,
$$
so that
\be\la{eqn:Sv1}
S = 2\pi i \Lambda  - \frac{2\pi^2 (k_R-k_L)}{\tilde{\mu}}~.
\ee
The second term vanishes when $k_R=k_L$ but otherwise not.
It represents a novel refinement when compared to analogous computations in higher dimensional AdS spaces.  

The individual entropy extremization conditions are
\begin{subequations}\la{eqn:ext3}\bea
\la{eqn:extwL}
\partial_{\tilde{\omega}_L} S &=& k_L \frac{2\tilde{\omega}_L}{\tilde{\mu}} + (2\Lambda - J_L) = 0 ~,\\
\la{eqn:extwR}
\partial_{\tilde{\omega}_R} S &=& - k_R \frac{2\tilde{\omega}_R}{\tilde{\mu}} -  J_R = 0 ~,\\
\la{eqn:extmu}
\partial_{\tilde{\mu}} S&=& - \frac{k_L \left( \tilde{\omega}_L^2 + \pi^2\right) - k_R \left( \tilde{\omega}_R^2 + \pi^2\right)}{\tilde{\mu}^2}  
- (\Lambda + P) = 0 ~.
\eea\end{subequations}
Using the constraint \eqref{eqn:indcond}, the first equation gives
\be \la{eqn:extmu2}
k_L \frac{\tilde{\mu} - 2\pi i }{\tilde{\mu} } = J_L -2 \Lambda
~~\Rightarrow~~
\frac{\pi i k_L}{\tilde{\mu}} = \Lambda - \frac{1}{2}(J_L-k_L) ~.
\ee
The entropy function therefore becomes 
\be\la{eqn:Sv2}
S = 2\pi i \left[ \Lambda  + \frac{i\pi}{\tilde{\mu}}(k_R - k_L) \right]
= 2\pi i\left[   \frac{k_R}{k_L} \Lambda - \frac{1}{2k_L} (k_R-k_L)(J_L-k_L) \right] \equiv 2 \pi i \Lambda_{\rm eff}~,
\ee
where we defined
\be\la{eqn:defLeff}
\Lambda_{\rm eff} = \frac{k_R}{k_L}  \Lambda - \frac{1}{2k_L} (k_R-k_L)(J_L-k_L) ~.
\ee
Rewriting the last extremization condition \eqref{eqn:extmu} using the others (\ref{eqn:extwL}-\ref{eqn:extwR})
and the expression for $\tilde{\mu}$ \eqref{eqn:extmu2} we find
\be\la{eqn:L}
 - \frac{1}{k_L} (\Lambda - \frac{1}{2}J_L)^2 + \frac{1}{4k_R} J_R^2 - (\Lambda + P)
 - \frac{1}{k^2_L} (k_R - k_L) (\Lambda - \frac{1}{2}(J_L - k_L))^2= 0 ~,
\ee
which we reorganize into a quadratic equation for $ \Lambda_{\rm eff}$:
\be\la{eqn:Leff}
 \Lambda^2_{\rm eff} - (J_L-k_L) \Lambda_{\rm eff}    + \frac{1}{4}(J_L - k_L)^2
+ k_R ( P +\frac{J_L}{2} - \frac{k_L}{4})   - \frac{1}{4} J_R^2   = 0 ~.
\ee
Selecting the root with negative imaginary part we find the extremized entropy function in terms of charges: 
\be\la{eqn:Sfinal}
S = 2\pi i \Lambda_{\rm eff} = 2 \pi \sqrt{k_R (P+\frac{J_L}{2} - \frac{k_L}{4}) - \frac{J_R^2}{4}} + \pi i (J_L-k_L)~.
\ee

For BPS black holes in higher dimensional AdS the standard prescription posits that charges must be constrained such that
the extremized entropy function is real \cite{Hosseini:2017mds,Choi:2018hmj}. Applying this rule in AdS$_3$ as well we find 
$$
J_L =  k_L ~,
$$
in agreement with the charge constraint \eqref{eqn:BPSJL} that we inferred from gravitational considerations.
After fixing the charges this way, the entropy function \eqref{eqn:Sfinal} is real with the value 
\be\la{eqn:Sv3}
S_{\rm BPS} = 2\pi \sqrt{ k_R ( P + \frac{1}{4}k_L) - \frac{1}{4} J_R^2}~, 
\ee
in agreement with the entropy \eqref{eqn:BPSS} of a BPS black holes in AdS$_3$ . 

In summary, in this subsection we applied the entropy extremization procedure to recover thermodynamic properties 
from the supersymmetric index \eqref{eqn:ind}. The computation is novel in that the index \eqref{eqn:ind} used here is more refined 
than the version \eqref{eqn:lnZindex2} that is directly analogous to higher dimensional cases, as explained at the 
end of subsection \ref{ssec:pfind}.


%
%



\subsection{Discussion: the Imaginary Part of the Entropy Function}\label{sec:imaginary}

The result of entropy extremization agrees with the gravitational side for the BPS black holes
in AdS$_3$ discussed here, as it does for their analogues in AdS$_{4, 5, 6, 7}$. 
However, in all these cases it is not entirely clear why the procedure works. 
In particular, it is somewhat mysterious how the reality condition on the entropy function gives the charge constraint obeyed by BPS black holes.
In this subsection we address this question in the AdS$_3$ context.

In order to understand the reality condition on the entropy function, recall how complex numbers enter in the first place.
We compute the supersymmetric index from the BPS partition function in \eqref{eqn:defI}, by imposing the complex constraint 
(\ref{eqn:indcond}) on the potentials:
\begin{eqnarray}
\label{eqn:inddef}
{\cal I} &=&\left. {\rm Tr}_{\rm BPS}~e^{\tilde{\mu}p +\tilde{\omega}_R j_R +\tilde{\omega}_L j_L} 
\right|_{\tilde\omega_L = \frac{\tilde\mu}{2} - i \pi} \nonumber \\
&=& {\rm Tr}_{\rm BPS} ~ e^{-i \pi j_L} e^{\tilde\mu (p +\frac{1}{2}j_L) + \tilde\omega_R j_R }~.
\end{eqnarray}
However, despite the appearance of a complex constraint, the index remains real as long as all potentials other than $\tilde\omega_L$ 
remain real, because the R-symmetry quantum number $j_L$ is quantized as an integer.
This, of course, is unsurprising since the complex number simply encodes the real grading $(-1)^F$.

Entropy extremization computes degeneracies (with negative signs for fermions) $d = e^S$
for states with specified quantum numbers from the index through a Legendre transform. 
Schematically for a system with one quantum number $j$ and chemical potential $\tilde\omega$ we have
\be
{\cal I} = \sum_j (-1)^{F(j)} d(j) \left( e^{\tilde\omega} \right)^j \quad \Leftrightarrow \quad
(-1)^{F(j)} d(j) = \oint \frac{d \tilde\omega}{2\pi i} e^{\log {\cal I} - \tilde\omega (j+1)} ~,
\ee
and entropy extremization amounts to computing the contour integral from a saddle point. 
However, this procedure does not introduce any genuinely complex numbers. We already noted that the index is real and the resulting degeneracies $d(j)$ must also be real, by definition.
Indeed, that is what our explicit result for the entropy function (\ref{eqn:Sfinal}) shows: although $\pi i (J_L - k_L)$ is complex, this term simply accounts for fermion statistics $e^{\pi i (J_L - k_L)}$
because $J_L - k_L \in \mathbb{Z}$. Thus the imaginary part of the entropy function has a perfectly acceptable physical interpretation and so there is no good reason {\it a priori} to demand that it vanish. 
It is puzzling, then, that the charge constraint required for regularity of the black hole geometry is precisely equivalent to reality of 
the entropy function. 

Our resolution of the puzzle is that the charge constraint originates from the BPS partition function which,
as we stressed in subsection \ref{ssec:pfind}, contains more information than the index.
However, due to a particular property of the BPS partition function (\ref{eqn:lnZBPS}),
the index inherits the data needed to infer the charge constraint.

To see this, consider the entropy function \eqref{eqn:Sfunction} that we extremized in subsection \ref{sec:entext}, written in terms of the BPS partition function:
\bea
S [ \tilde\mu, \tilde\omega_{R} ] &=&
\left( \ln Z_{\rm BPS} - \tilde\mu P - \tilde\omega_R J_R - \tilde\omega_L J_L\right) |_{\tilde\omega_L = \frac{\tilde\mu}{2} - i \pi}  \nonumber ~.
\eea
Here we explicitly substitute $\tilde\omega_L = \frac{\tilde\mu}{2} - i \pi$, rather than employing a Lagrange multiplier. Also, we omitted the
supersymmetric Casimir energy for clarity, as it is immaterial to our argument. Extremization of the entropy function over $\tilde{\mu}$ gives
%
\bea
\frac{d}{d\tilde\mu}(\ln Z_{\rm BPS})|_{\tilde\omega_L = \frac{\tilde\mu}{2} -i \pi}
&=& \frac{\partial}{\partial \tilde\mu} \ln Z_{\rm BPS} + \frac12 \frac{\partial}{\partial \tilde\omega_L}\ln Z_{\rm BPS} = P + \frac{1}{2}J_L~. \label{eqn:PandJL} 
\eea
%
This reproduces the standard formulae for macroscopic charges $P$ and $J_L$ in the canonical ensemble,
but only for the combination $P+\frac{1}{2}J_L$.
The outcome that only one combination of $P$ and $J_L$ appears is expected because, as seen in (\ref{eqn:inddef}), 
the index does not distinguish the two charges $P$, $J_L$, it only depends on their combination $P + \frac{1}{2}J_L$.
However, we found in \eqref{eqn:BPSJL2} that the charge constraint $J_L = k_L$
originates from averaging over the $j_L$ quantum number alone. In other words,
the charge constraint follows from separating (\ref{eqn:PandJL}) into two independent equations,
one for $P$ and another for $J$, a step that is usually not justified. 

However, the situation at hand is special, because the BPS partition function
(\ref{eqn:lnZBPS}), and so the entropy function $S= \ln Z_{\rm BPS} - \tilde\mu P - \tilde\omega_R J_R - \tilde\omega_L J_L$, 
are linear functions of $\tilde\omega_L$, and also real functions of all other potentials.
Therefore, provided that $\tilde\mu$ and $\tilde\omega_R$ are real and $\tilde\omega_L = \frac{\tilde\mu}{2}-i \pi$
is the only source of complex numbers,
\bea
\text{Im}(\ln Z_{\rm BPS})
&=& (\text{Im}~\tilde\omega_L) \cdot \frac{\partial}{\partial \tilde\omega_L} \ln Z_{\rm BPS}~,
\nonumber \\
\Rightarrow~ \text{Im}~S
&=& (\text{Im}~\tilde\omega_L) \cdot \left( \frac{\partial}{\partial \tilde\omega_L} \ln Z_{\rm BPS} - J_L\right)~.
\eea
The requirement that $S$ be real gives
\be
 \text{Im}~S=0 \quad \Leftrightarrow \quad \frac{\partial}{\partial \tilde\omega_L} \ln Z_{\rm BPS} - J_L =0~,
\ee
which becomes the charge constraint $J_L = k_L$.
This is how, upon introduction of complex numbers via $\tilde\omega_L = \frac{\tilde\mu}{2}-i \pi$,
reality of the entropy function mimics extremization with respect to a potential that is an independent variable 
only in the BPS partition function and not in the index.

To summarize, the BPS charge constraint $J_L=k_L$ is a piece of information that is
contained in the partition function (\ref{eqn:lnZBPS}) but not in the index (\ref{eqn:ind}), 
because dependence on two potentials $\tilde\mu$ and $\tilde\omega_L$ are lumped together in the index.
It is only because the BPS partition function is i) a real function of all potentials and ii)
linear in $\tilde\omega_L$,  that the dependence on $\tilde\omega_L$ alone can be extracted from the index,
as it is encoded in the imaginary part. Were it not for these features, a principled derivation of the charge constraint would follow only from
the BPS partition function and not from the index, which depends on one fugacity less.

It is unclear if the analogous mechanism applies to asymptotically AdS$_5$ BPS black holes where, in fact, 
the correct charge constraint can be derived by demanding reality of 
the entropy function. Recent progress on the superconformal index of the dual $\mathcal{N}=4$ Super-Yang-Mills theory
relies heavily on the \emph{modified} index \cite{Choi:2018hmj,Kim:2019yrz,Copetti:2020dil}
where the role of $(-1)^F$ is played by $e^{i\pi r}$ with $r$ the $U(1)$ R-charge of 4D $\mathcal{N}=1$ theory.
The modified index is a Witten index that counts only $\frac{1}{16}$-BPS states
and exhibits deconfined behavior for some complex phases of the fugacities.
However, it is no longer a manifestly real function even when the fugacities are real,
because $r$ is not integer-quantized. In this situation reality of the extremized entropy function is not an \emph{a priori}  
principled way to extract additional information from the index.


\subsection{Potentials and the BTZ nAttractor Mechanism} 

The entropy function is constructed from the index, yet it encodes data characterizing black holes that are
not even BPS. In this subsection we illustrate this claim and, in the process, develop a spacetime interpretation of the potentials that extremize the entropy function, following analogous 
computations for black holes in higher dimensional AdS spaces \cite{Larsen:2019oll,Larsen:2020lhg}. 

The value of the potential for 3D angular momentum at the extremum of the entropy function was determined in \eqref{eqn:extmu2}: 
\begin{subequations}
\be
\label{eqn:munads}
\tilde{\mu}  = \frac{2\pi i k_L}{ 2\Lambda - (J_L-k_L)} = -\frac{\pi k_R}{\sqrt{ k_R ( P + \frac{1}{4}k_L) - \frac{1}{4} J_R^2}} ~. 
\ee
In the second equation we first take $J_L=k_L$ satisfied by BPS black holes
and then the denominator $\Lambda$ becomes purely imaginary with value given implicitly in \eqref{eqn:L}.
We choose its sign consistently with \eqref{eqn:Sfinal} and with $\tilde\mu < 0$.
The constraint \eqref{eqn:indcond} and the extremization condition on $\tilde{\omega}_R$ \eqref{eqn:extwR} then easily give
\bea
\label{eqn:wLnads}
\tilde{\omega}_L &=& \frac{1}{2} \left( \tilde{\mu} - 2\pi i \right) = - \frac{\pi k_R}{2\sqrt{ k_R ( P + \frac{1}{4}k_L) - \frac{1}{4} J_R^2}}  - i \pi ~,
\\
\label{eqn:wRnads}
\tilde{\omega}_R &=& - \frac{J_R}{2k_R} \tilde{\mu}= \frac{\pi J_R}{2\sqrt{ k_R ( P + \frac{1}{4}k_L) - \frac{1}{4} J_R^2}}
~.
\eea\end{subequations}
These potentials are real, except for the imaginary part of $\tilde{\omega}_L$ which implements the boundary
condition needed for the index. They are derived from the index, an object protected by supersymmetry,
yet their real parts can be identified with physical potentials in spacetime \cite{Hosseini:2017mds}.
More precisely, they correspond to features of the potentials that {\it break} supersymmetry. 

In order to establish this we adapt the nearAdS attractor mechanism known in higher dimensions
\cite{Larsen:2018iou,Hong:2019tsx} to BTZ black holes \cite{Banados:1992wn,Banados:1992gq}. 
Accordingly, consider a general asymptotically AdS$_3$ geometry of the form 
\be
\label{eqn:gen3Dgeom}
ds^2 = - \frac{r^4 - r^4_0}{\ell^2 R^2(r)} dt^2 + \frac{\ell^2 r^2}{r^4 - r^4_0} dr^2 + R^2(r) \left( d\phi + \frac{\mu(r)}{\ell} dt \right)^2 ~, 
\ee
where the function $R^2(r) \sim r^2$ for large $r$ to ensure the correct asymptotics. The BTZ black hole at hand is the special case where 
$r^2_0 = \frac{1}{2} (r^2_+ - r^2_-)$ and the functions specifying the geometry
 are \footnote{The standard radial coordinate for the BTZ black hole is $r^2_{\rm BTZ} = R^2_{\rm here}$. The shifted radial coordinate here is 
a close analogue of the radial coordinate that is appropriate in higher dimensional cases.}
\bea
\label{eqn:R2(r)}
R^2(r) &=& r^2  + \frac{1}{2} ( r^2_+ + r^2_-)~,\cr
\label{eqn:mu(r)}
\mu(r) & = &  \frac{r_+ r_-}{R^2(r)} ~,
\eea
in terms of the parameters $r^2_\pm$ that are related to physical black hole variables as
\bea
M &=& \frac{r^2_+ + r^2_-}{8G_3\ell^2}~,
\cr
P &= & \frac{r_+ r_-}{4G_3\ell} ~.
\eea
We denote 3D angular momentum by $P$ to conform with notation elsewhere in this article. 

Regularity of the Euclidean geometry at the horizon $r^2 = r^2_0$ determines the temperature of any black hole of the 
form \eqref{eqn:gen3Dgeom} as
$$
T =  \frac{r^2_0}{\pi\ell^2 R(r_0)}~.
$$
In the extremal case $r_0^2=0$ and the inner and outer horizons coincide at $r^2=0$, but at non-zero temperature they move to $\pm r^2_0$, respectively. The associated entropy change is entirely captured by the increase in ``area" due to the event horizon moving outwards by $\Delta r^2 = r^2_0$: 
$$
\Delta S =  \left. \frac{1} {4G_3} \cdot 2\pi\partial_{r^2} R\right|_{r^2 = r^2_0} \Delta r^2 =  
\left. \frac{\pi^2\ell^2} {4G_3} ~  \partial_{r^2} R^2\right|_{r^2 = r^2_0}    T~.
$$
The BTZ black hole \eqref{eqn:R2(r)} has $\partial_{r^2} R^2=1$ so the near extremal heat capacity
is linear in temperature $C_T\sim T$ with constant of proportionality 
\begin{equation}
\label{eqn:nattCT}
\frac{C_T}{T} = \frac{\pi^2\ell^2} {4G_3} = \pi^2 k_L \ell ~,
\end{equation}
where we used the Brown-Henneaux formula $ \frac{\ell}{4G_3} = k_L $ \cite{Brown:1986nw}
for excitations of a BPS black hole with its $L$-sector in the ground state.
\footnote{The refinements needed to distinguish between $k_L$ and $k_R$ in AdS$_3$ were discussed in \cite{Kraus:2005zm}.} 

Similarly, the dimensionless 3D rotational velocity \eqref{eqn:mu(r)} is $\mu (0) = 1$ for the BPS black hole where $R^2 = r^2_+=r^2_-$. For a
nearBPS black hole it is changed by 
\be
\Delta \mu = - \left. \frac{r_+r_-}{R^4}
\partial_{r^2} R^2\right|_{r^2 = r^2_0} \Delta r^2 =  - \frac{\pi\ell^2}{R} ~T ~.
\ee
This contribution is negative because the nearBPS rotational velocity is below the speed of light.
The ``area" of the event horizon is $2\pi R$ so we can rewrite the rescaled potential \eqref{eqn:BPSlimit}
in terms of the BPS entropy and find
\begin{equation}
\label{eqn:nattrmu}
\tilde{\mu}  = \frac{\Delta\mu}{T} = -  \frac{2\pi^2 \ell k_R}{S_{\rm BPS}} 
~.
\end{equation}
We used the Brown-Henneaux formula $ \frac{\ell}{4G_3} = k_R$ for BPS states preserving the $L$-sector ground state. 
The result agrees in the unit $\ell=1$ with \eqref{eqn:munads} from entropy extremization,
given \eqref{eqn:BPSS}, as expected. 

We defined both the specific heat \eqref{eqn:nattCT} and the nearBPS rotational velocity \eqref{eqn:nattrmu} as response coefficients for
the black hole becoming near-extremal, by adding a small temperature. However, the computation in this subsection shows that we can equally interpret these parameters as characterizing the BPS black hole, albeit slightly away from its event horizon. This is the situation described
in low energy effective field theory by the nAdS$_2$/nCFT$_1$ correspondence and seems like the most appropriate for discussions of the index.


\subsection{The Hawking-Page Transition for BPS Black Holes} 

The thermodynamics of black holes in AdS spacetimes sheds light on the phase diagram of gauge theories (and their relatives) at strong 
coupling \cite{Witten:1998zw,Aharony:2003sx}. This relation is interesting even for BPS black holes described by an index, despite the protection against phase transitions due to supersymmetry. For example, interpreting the index as a conventional free energy gives, for
BPS black holes in AdS$_5$, a phase diagram that is surprisingly similar to that of the Schwarzschild-AdS$_5$ black hole \cite{Choi:2018vbz,Copetti:2020dil}. In this subsection we give a perspective on such higher dimensional BPS phase diagrams by discussing their analogue in AdS$_3$. 

The BPS partition function \eqref{eqn:lnZBPS} gives the free energy in the BPS limit as
\begin{equation}
\label{eqn:ZBPSfreeen}
{\cal W} =  - {\rm ln} Z_{\rm BPS}  = - \frac{1}{2} k_L  \beta  -
k_L  \tilde{\omega}_L + \frac{k_R}{\tilde{\mu}} \left( \pi^2 + \tilde{\omega}^2_R\right)  + \frac{1}{4} k_L \tilde{\mu}  ~.
\end{equation}
We define the BPS free energy without the factor $1/\beta$ appearing in standard thermodynamics. 
Local thermodynamic stability can be probed by the compressibility matrix
\be\la{eqn:defcomp}
K^{ij} = - \left(\frac{\partial^2 {\cal W}}{\partial\Phi_i \partial\Phi_j}\right)_T~,
\ee
where $\{ \Phi_i\} = \{ \tilde\mu,\tilde\omega_R, \tilde\omega_L \}$ collectively refer to the potentials. The
potential $\tilde\omega_L$ parametrizes a direction that decouples and is entirely flat.
The remaining two directions are spanned by $\tilde{\omega}_R$ and $\tilde{\mu}$,
and the free energy has response coefficients 
\be
- \begin{pmatrix}
\dfrac{\partial^2 {\cal W}}{\partial\tilde{\mu}^2} & \dfrac{\partial^2 {\cal W}}{\partial\tilde{\mu}\partial\tilde{\omega}_R}  \\
 \dfrac{\partial^2 {\cal W}}{\partial\tilde{\omega}_R\partial\tilde{\mu}}  & \dfrac{\partial^2 {\cal W}}{\partial\tilde{\omega}^2_R}
\end{pmatrix}
=
\begin{pmatrix}
 - \dfrac{2k_R}{\tilde{\mu}^3}(\pi^2 + \tilde{\omega}^2_R) &  \dfrac{2k_R\tilde\omega_R}{\tilde\mu^2}   \\
 \dfrac{2k_R\tilde\omega_R}{\tilde\mu^2}  &  - \dfrac{2k_R}{\tilde{\mu}} 
\end{pmatrix}~.
\ee
Recalling that $\tilde{\mu}<0$, both eigenvalues of the matrix are positive. Therefore, the compressibility matrix is positive definite and
the system is locally stable. 

The formula \eqref{eqn:ZBPSfreeen} expresses the standard Cardy asymptotics of CFT$_2$ but in a notation that is adapted for comparison with 
BPS black holes in higher dimensional AdS. The linear-in-$\beta$ term encodes the supersymmetric Casimir energy \eqref{eqn:SUSYCast}. 
Similarly, the linear-in-$\tilde{\omega}_L$ term encodes the charge constraint $J_L = k_L$ \eqref{eqn:BPSJL2}.
Both of these linear contributions depend only on $k_L$ so they are properties of the theory rather than the state.
They can be removed without losing any physical information, by Legendre transform to a microcanonical ensemble
that fixes the charges $E$ and $J_L$ rather than the potentials $\beta$ and $\tilde{\omega}_L$.
This feature shows that a linear shift in the potentials $\beta$, $\tilde{\omega}_L$ is inconsequential
so we can remove the first two terms in \eqref{eqn:ZBPSfreeen} entirely, not even a constant is left behind. 

The remaining two terms in \eqref{eqn:ZBPSfreeen} are negative because $\tilde{\mu}$ is required to be negative,
as discussed above \eqref{eqn:extlimit}. Apart from the sign, the potential $\tilde\mu$ can be interpreted as an inverse ``temperature"
\begin{equation}
\label{eqn:Teffdef}
T_{\rm eff} =  - \tilde{\mu}^{-1}~. 
\end{equation}
The physical temperature vanishes, as always for BPS states, but this effective BPS temperature expresses the
usual physical intuition that a large value corresponds to large occupancy numbers.
The sum of the two ``thermal" contributions to the free energy \eqref{eqn:ZBPSfreeen} are bounded from above 
$$
{\cal W} \leq {\cal W}_{\rm max} = - \pi \sqrt{k_R k_L}~,
$$
with equality when $\tilde{\omega}_R=0$ and 
\begin{equation}
\label{eqn:muHP}
\tilde{\mu} = \tilde{\mu}_{\rm HP} = - 2\pi \sqrt{\frac{k_R}{k_L}}~.
\end{equation}
The index ``HP" anticipates that we shortly interpret the special value \eqref{eqn:muHP} as the Hawking-Page transition temperature. 

The standard modular S-transformation takes $\tilde{\mu} \to \frac{4\pi^2}{\tilde{\mu}}$ and, at least at a first glance, the free energy \eqref{eqn:ZBPSfreeen} suggests that such a high/low temperature duality could persist in the effective description, perhaps inherited from an underlying $SL(2,\mathbb{Z})$ symmetry and subject to the interesting refinement that the self-dual point would have to be rescaled from $2\pi$ to $2\pi \sqrt{\frac{k_R}{k_L}}$. Unfortunately, as we explain next, this suggestion does not hold up to closer scrutiny.

In bulk AdS$_3$ quantum gravity, 
modular transformation interchanges the high temperature black hole phase where (Euclidean) temperature is contractible
with the low temperature AdS gas phase where it is the spatial circle that is contractible.
Indeed, in the complete CFT$_2$ there are infinitely many saddle points related by $SL(2,\mathbb{Z})$ symmetry,
corresponding to the thermal gas and a family of black hole images \cite{Dijkgraaf:2000fq}. However, the free energy
\eqref{eqn:genlnZ} that we study throughout this paper does not represent a complete CFT$_2$, 
it is just the classical contribution from a single saddle point, that of the simplest black hole. 
It is related to the thermal gas saddle point by the $SL(2,\mathbb{Z})$ symmetry in the full theory, but the map is nontrivial. 
Duality takes $\tilde{\omega}_R \to \frac{2 \pi i \tilde{\omega}_R}{\tilde{\mu}}$, 
flipping the sign of the term in \eqref{eqn:ZBPSfreeen} that is proportional to $\tilde{\omega}^2_R$. Moreover, the free energy is not invariant, its transformation adds a term proportional to 
$\tilde{\omega}^2_R$ such that no term of this form remains, and it adds yet another term proportional to $\tilde{\omega}^2_L$. In this way, 
the underlying high/low temperature duality relates the black hole and the thermal gas while also exchanging the $L$ and $R$ sectors of the CFT$_2$. We expect that similar mechanisms are possible in higher dimensions.

The procedure followed when analyzing BPS black holes in higher dimensional AdS spaces
suggests yet another perspective on the free energy \eqref{eqn:ZBPSfreeen}.
Motivated by the supersymmetric index \eqref{eqn:defI}, we cancel the linear-in-$\beta$ term that gives the supersymmetric Casimir energy but we then evaluate the linear-in-$\tilde{\omega}_L$ term by imposing the
constraint \eqref{eqn:indcond}. This gives the index-inspired free energy
\begin{equation}
\label{eqn:ZBPSfreee}
{\cal W}_I =  \frac{k_R}{\tilde{\mu}} \left( \pi^2 + \tilde{\omega}^2_R\right)  - \frac{1}{4} k_L \tilde{\mu}
=  - k_R \left( \pi^2 + \tilde{\omega}^2_R\right)T_{\rm eff}  + \frac{k_L}{4T_{\rm eff}} ~,
\end{equation}
that is an AdS$_3$ analogue of the free energy taken as a basis for discussions of the confinement/deconfinement
transition for black holes in higher dimensional AdS \cite{Choi:2018vbz,Copetti:2020dil}. 
Note that the second term $ \frac{k_L}{4T_{\rm eff}}$ now gives a positive contribution to the free energy.

\begin{figure}
\begin{center}
\includegraphics[width=6cm]{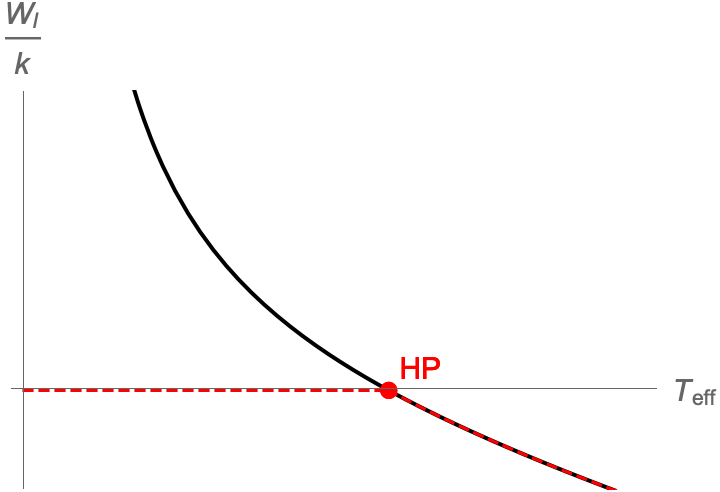}
\caption{\label{fig:HP} Index-inspired free energy \eqref{eqn:ZBPSfreee} as a function of effective temperature
$T_{\rm eff}= - \tilde\mu^{-1}$, not drawn to scale. The red dashed line represents the thermal gas phase
for lower temperature, and the large black hole phase for higher temperature.
The red dot represents the Hawking-Page transition point between the two.}
\end{center}
\end{figure}

The index-inspired free energy \eqref{eqn:ZBPSfreee} is plotted, in units of $k$,
as a function of effective temperature $T_{\rm eff}$ in Figure \ref{fig:HP}. We interpret the phase diagram
in analogy with the AdS-Schwarzschild case and discussions of BPS black holes in higher dimensional AdS.
The high temperature phase where $\mathcal{W}_I<0$ is the black hole phase, or more precisely the ``large" black hole phase.
At lower temperature, the part of the line where $\mathcal{W}_I>0$ is the ``small" black hole phase.
This phase is unstable because there is an entirely different saddle point, not captured by the free energy formula
we analyze, that corresponds to the thermal gas with no black hole and has $\mathcal{W}_I=0$ at all temperatures.
The Hawking-Page transition point is where the line crosses $\mathcal{W}_I=0$, at the temperature 
corresponding to \eqref{eqn:muHP}.

The index-inspired free energy assigns the entire expression \eqref{eqn:ZBPSfreee} to the black hole while the BPS free energy \eqref{eqn:ZBPSfreeen}
interprets the two last terms in \eqref{eqn:ZBPSfreeen} as the black hole and thermal contributions, respectively. The two approaches
therefore differ physically, but they give the same transition temperature, because acting with $\left. -\tilde{\mu}\partial_{\tilde{\mu}}\right|_{\tilde{\omega}_L}$ on the free energy \eqref{eqn:ZBPSfreeen} is exactly equivalent to imposing the real part of the constraint \eqref{eqn:indcond}. As in subsection \ref{sec:imaginary} this is possible because the free energy depends linearly on $\tilde{\omega}_L$.

\section{NearBPS Black Holes}\la{sec:nBPS}

In this section we generalize the description of AdS$_3$ BPS black holes discussed in the previous sections
and study the thermodynamics of small deviations away from the BPS limit. This adapts to AdS$_3$ the 
nearBPS black hole thermodynamics in AdS$_{4, 5, 7}$ that was studied in \cite{Larsen:2019oll,Larsen:2020lhg}. 
The simplifications in AdS$_3$ clarify their higher dimensional analogues. 
 
\subsection{Introducing NearBPS Thermodynamics}\la{sec:nBPSthermo}

We first evaluate the macroscopic quantum numbers for AdS$_3$ black holes slightly away from the BPS limit. The organizing principle,  
stressed in subsection \ref{subsec:codim2}, is that BPS black holes are co-dimension two in parameter space.
The two conditions satisfied by BPS black holes were presented, by the thermodynamic interpretation in subsection \ref{sec:BPSthermo}, 
as extremality $T=0$ and, in addition, the vanishing of
the potential $\varphi = 1 + \mu - 2\omega_L $ introduced in \eqref{eqn:BPSpotentials}.
Therefore, the nearBPS regime is characterized by $T$ and $\varphi$ that are small but not necessarily 
zero.\begin{footnote}{
The deviations $T$ and $\varphi$ need not be small, the general partition function \eqref{eqn:genlnZ} is valid for any non-BPS black hole. Taking them small illuminates the relation between the BPS and nearBPS regimes. 
Additionally, considerations for small $T$ and $\varphi$ are direct analogues of discussions of black holes in higher dimensional AdS spaces.
}\end{footnote}
We take the two parameters $T$ and $\varphi$ to be of the same order in smallness:
$$T \sim \varphi \sim \epsilon \ll 1~.$$

In the canonical ensemble, the four macroscopic charges of a generic nonBPS black hole (\ref{eqn:Pgen}-\ref{eqn:Egen}) 
are functions of four independent conjugate potentials. We can pick a basis where the potentials are 
$\tilde\mu = \beta (\mu-1)$, $\tilde\omega_R =\beta\omega_R$, $T$, and $\varphi$. 
For given values of $\tilde\mu$ and $\tilde\omega_R$, we now expand (\ref{eqn:Pgen}-\ref{eqn:JRLgen})
to linear order in $T$, $\varphi$ and find
\begin{subequations}\la{eqn:chargenearBPS}\bea
\label{eqn:PnearBPS}
P &=&
\frac{k_R}{\tilde{\mu}^2} (\pi^2 + \tilde{\omega}^2_R)-\frac{k_L}{4} + \frac{k_L}{4} \varphi + \ldots
= P_* + \frac{k_L}{4} \varphi + \ldots 
~,\\
J_{L} & = &  k_L - \frac{k_L}{2} \varphi + \ldots 
= {J_L}_* - \frac{k_L}{2}  \varphi +\ldots 
~, \\
\label{eqn:JRnearBPS}
J_{R} & = &  - \frac{2k_{R}}{\tilde{\mu}}  \tilde{\omega}_{R} = {J_R}_*~. 
\eea\end{subequations}
The dots denote terms of order $\mathcal{O}(\epsilon^2)$ that we neglect. The quantities with an asterisk refer to the
values of the charges (\ref{eqn:BPScharge}-\ref{eqn:BPSJL}) in the strict BPS limit where $T=0$ and $\varphi=0$. The formulae show
that, in our basis of potentials, none of the charges depend on temperature $T$ to linear order, and 
$J_R$ depends on neither $T$ nor $\varphi$ to any order. The potential $\varphi$ is a source for the charges but leaves fixed the combination $P + \frac{1}{2}J_L$ that the index is sensitive to. 

We also want to expand the energy \eqref{eqn:Egen} in $T$ and $\varphi$.
However, recall that, for given charges $P$ and $J_L$, the energy is bounded from
below by $E_{\rm BPS} = P + J_L -\frac12 k_L$.
Therefore, rather than computing the energy by itself, it is instructive to expand the excitation energy $E - E_{\rm BPS}$ above the BPS 
bound. \begin{footnote}{Because we are also considering $J_L$ away from its BPS limit $k_L$,
the ``BPS" energy $E_{\rm BPS}$ is not necessarily the energy of a BPS black hole.
It is the energy of a hypothetical ``black hole" that is supersymmetric but not
necessarily regular, for given charges.}\end{footnote}
It vanishes at linear order but at quadratic order we find
\begin{eqnarray}\label{eqn:EnearBPS}
E - E_{\rm BPS} &=& 
\frac{2k_L}{\beta^2(1+\mu)^2} (\pi^2 + \beta^2 \omega^2_L) - \frac{2k_L\omega_L}{1 + \mu}
+\frac12 k_L
\nonumber \\
&=&
\frac{1}{8} k_L\Big( (2\pi T)^2+ \varphi^2  \Big)~.
\end{eqnarray}

The formulae (\ref{eqn:chargenearBPS}-\ref{eqn:EnearBPS}) characterize the low lying excitations of
a BPS black hole which, by definition, is both extremal and supersymmetric. This ground state has the smallest possible mass for its 
charges and, to preserve supersymmetry, the charges are constrained by $J_L=k_L$.
The formulae make explicit that these two conditions correspond to two orthogonal directions that violate BPS-ness of the black hole:
\begin{itemize}
\item
One direction raises the temperature, so that the mass increases by $\frac18 k_L (2 \pi T)^2$
while charges remain unchanged. Conversely, as noted after \eqref{eqn:chargenearBPS}, all charges 
are independent of $T$.
\item
Another direction turns on the potential $\varphi$ while maintaining zero temperature.
As a result, the charges \eqref{eqn:chargenearBPS} are shifted by terms that are linear in $\varphi$.
The energy of the resulting extremal but non-supersymmetric black hole is given by \eqref{eqn:Eext},
which is higher than $E_{\rm BPS}$ by
\bea\la{eqn:Eext-EBPS}
E_{\rm ext} - E_{\rm BPS} &=& \left( P + \frac{J_L^2}{2k_L} \right) - \left( P+J_L-\frac12 k_L \right)
\nonumber \\
&=& \frac{1}{2k_L} (J_L-k_L)^2
\nonumber \\
&=& \frac{1}{8} k_L \varphi^2~,
\eea
in agreement with \eqref{eqn:EnearBPS}.
\end{itemize}


Expanding the entropy \eqref{eqn:Sgen} at linear order 
in $T$ and $\varphi$ gives
\begin{eqnarray}
\label{eqn:SnearBPS}
S & = &   - \frac{2k_R \pi^2}{\tilde{\mu}}  + \pi^2 k_L T +\ldots = S_* + \pi^2 k_L T  + \ldots ~.
\end{eqnarray}
The entropy has no term that is linear in $\varphi$, but only a term that is linear in $T$.
This term indicates a heat capacity $C_T$ that is linear in temperature with a value
\be\la{eqn:CT}
\frac{C_T}{T} =  \pi^2 k_L~.
\ee
This coefficient, computed from black hole thermodynamics, agrees with the result of the nAttractor mechanism
\eqref{eqn:nattCT} in the unit $\ell=1$, which is derived directly from the geometry of the supersymmetric black hole. 

In the expression for the excitation energy \eqref{eqn:EnearBPS}, the heat capacity enters as a term that is quadratic in the temperature $T$.  Furthermore, drawing analogy between the potential $\frac{\varphi}{2\pi}$ and an electric potential, we interpret the 
coefficient of the term quadratic in $\varphi$ as the capacitance. The energy formula shows that these two linear response coefficients 
are identical, up to possible differences in notation and terminology. We introduce a parameter $C_{\varphi}$ in lieu of capacitance, 
in order to stress this fact:
\be\la{eqn:CT=Cphi}
\frac{C_T}{T} = \frac{C_\varphi}{T} =  \pi^2 k_L ~.
\ee
This agreement is a nontrivial consequence of ${\cal N}=2$ supersymmetry. For example, it is built into the ${\cal N}=2$ superschwarzian description of the low energy excitations, i.e. the nAdS$_2$/nCFT$_1$ correspondence.

\subsection{The First Law of NearBPS Thermodynamics}\la{sec:1stLawnBPS}

As a check on our computations and our understanding, we can now explicitly verify the first law of thermodynamics
\begin{eqnarray}
\label{eqn:firstlawlin}
T dS &=& dE- \mu dP  - \omega_R dJ_R - \omega_L dJ_L \nonumber \\
&=& d(E-P - J_L) - (\mu-1) dP  - \omega_R dJ_R - (\omega_L - 1) dJ_L ~,
\end{eqnarray} 
in the nearBPS regime. For variations within the BPS surface, ${J_L}_*= k_L$ is constant so $dJ_L=0$, and
$d(E-P-J_L) = 0$ follows from $E_*= P_* + {J_L}_* - \frac12 k_L$ because $k_L$ is constant.
Therefore, the first law within the BPS surface reduces to:
\begin{equation}
\label{eqn:BPSsurface}
dS_* = - \tilde{\mu} dP_* - \tilde{\omega}_R d{J_R}_*~.
\end{equation}
%
This is indeed satisfied by the BPS expressions (\ref{eqn:BPScharge}-\ref{eqn:BPSS}): the variables $S_*$, $P_*$, and ${J_R}_*$ depend on the potentials $\tilde{\omega}_R$ and $\tilde{\mu}$ only, and in such a way that the linear relation (\ref{eqn:BPSsurface}) is satisfied. Thus (\ref{eqn:BPSsurface}) parametrizes the 2D surface of BPS black holes. 

Taking into account the BPS surface (\ref{eqn:BPSsurface}), we can rewrite the more general first law (\ref{eqn:firstlawlin})
as an equation for excitations above the BPS surface:
\be\label{eqn:firstlawnBPS}
T d(S-S_*) = d(E-P - J_L ) - (\mu-1) d(P-P_*) - (\omega_L - 1) dJ_L ~.
\ee
There is no differential $dJ_R$ because ${J_R}_*={J_R}$. Variations of $J_R$ do not influence the excitations,
they correspond to motion entirely within the BPS surface. We now use \eqref{eqn:chargenearBPS}
to evaluate two of the terms on the right hand side: 
\be\la{eqn:firstlawnBPS2}
(\mu-1) d(P-P_*)  + (\omega_L - 1)dJ_L
= \frac{1}{4}k_L [ (\mu-1) - 2(\omega_L - 1)] d\varphi = \frac{1}{4}k_L\varphi d\varphi ~.
\ee
At this point we can verify that \eqref{eqn:EnearBPS} and \eqref{eqn:SnearBPS} satisfy the first law for excitations above the BPS surface \eqref{eqn:firstlawnBPS}:
$$
T d\underbrace{(S-S_*)}_{\pi^2 k_L T} = 
d\underbrace{(E-P - J_L )}_{ \frac{1}{2}\pi^2 k_LT^2 + \frac{1}{8} k_L\varphi^2}- 
\underbrace{[ (\mu-1) d(P-P_*) + (\omega_L - 1) dJ_L]}_{\frac{1}{4}k_L\varphi d\varphi}  ~.
$$

\subsection{NearBPS Thermodynamics in the Canonical Ensemble}

On the gravitational side of the AdS/CFT correspondence it is natural to study thermodynamics in the canonical ensemble, with 
potentials specified and the conjugate charges incorporated as subsidiary variables. In this subsection we first discuss the nearBPS potentials and 
then the nearBPS free energy. 

Inverting the relations \eqref{eqn:PnearBPS} and \eqref{eqn:JRnearBPS} between $(P, J_R)$ and $(\tilde\mu , \tilde\omega_R)$ 
we find
\begin{subequations}\la{eqn:tildepotnearBPS}
\bea
\tilde\mu &=& \frac{\mu - \mu_*}{T}  = - \frac{\pi k_R}{\sqrt{k_R(P + \frac{1}{4}k_L) - \frac{J_R^2}{4}}} + O(\epsilon)~, \\
\tilde\omega_R &=& \frac{\omega_R-{\omega_R}_*}{T}  =  \frac{\pi J_R}{2\sqrt{k_R(P + \frac{1}{4} k_L)- \frac{J_R^2}{4}} }+ O(\epsilon)~,
\eea
where the BPS values of the potentials are $\mu_* = 1$, ${\omega_R}_*=0$ and the sign for the square root was chosen so $\tilde\mu < 0$.
The nearBPS corrections of order $\epsilon \sim T \sim \varphi$ are not needed.
Therefore, at this order, the equations are essentially the same as the BPS relations \eqref{eqn:BPScharge},
and they also agree with the real part of the potentials (\ref{eqn:munads}, \ref{eqn:wRnads})
determined by extremization of the BPS entropy function, and with the value \eqref{eqn:nattrmu} from the spacetime solution. 
However, in the nearBPS thermodynamics,
terms of $O(\epsilon)$ are merely small, the strict limit $\epsilon\to 0$ is not implemented.
This distinction is helpful when computing the analogous formula for $\tilde\omega_L$,
using the definition of $\varphi$ \eqref{eqn:BPSpotentials}
\bea\la{eqn:nearBPS}
\tilde\omega_L &=& \frac{\omega_L-{\omega_L}_*}{T} = - \frac{\pi k_R}{2\sqrt{k_R( P + \frac{1}{4} ) k_L- \frac{J_R^2}{4}}} -\frac{\varphi}{2\pi T} \cdot \pi + O(\epsilon)~.
\eea
\end{subequations}
Here ${\omega_L}_* = 1$ and we used ${\omega_L}<1$ to determine the sign for the square root.
The potentials $\varphi$ and $T$ both vanish in the strict BPS limit $\epsilon\to 0$
but {\it a priori} the ratio $\frac{\varphi}{2\pi T}$ can take any value without obstructing BPS saturation. 

The index corresponds to an analytical continuation of the black hole that takes $\frac{\varphi}{2\pi T}\to i $, as one can see from \eqref{eqn:indcond}.
In this sense the real and imaginary parts of the result \eqref{eqn:nearBPS} for the potential $\tilde\omega_L$ both coincide
with the complex value \eqref{eqn:wLnads} that was derived by extremization of the entropy function. The agreement between 
imaginary parts is not very impressive in AdS$_3$ because it is very simple, $\tilde\mu$ and $\tilde\omega_R$ are both independent of 
$\frac{\varphi}{2\pi T}$. However, analogous agreements persist in higher dimensional AdS where they are more elaborate, with multiple potentials involved \cite{Larsen:2019oll,Larsen:2020lhg}. 

In the canonical ensemble all thermodynamic data --- charges, energy, entropy --- is contained in Gibbs' free energy
\be\la{eqn:defG}
G \equiv -\frac{1}{\beta} \ln Z~.
\ee
In the nearBPS regime where we expand in small $T$, $\varphi$ for given $\tilde\mu$, $\tilde\omega_R$,
\bea
\label{eqn:GnearBPS}
G &=& -\frac{1}{\beta} \left( -\frac{k_R}{\tilde{\mu}} \Big( \pi^2 + \tilde{\omega}_R^2 \Big) 
+ \frac{k_L}{\tilde{\mu}+2\beta} \Big( \pi^2 + (\tilde{\omega}_L+\beta)^2\Big)  \right)
\nonumber \\
&=&
G_{\rm BPS} - \frac{k_L}{8} \Big( \varphi^2 +(2\pi T)^2 \Big) + \ldots~,
\eea
up to quadratic order in $T$ and $\varphi$, and we have 
\bea\la{eqn:GBPS}
G_{\rm BPS} &=&
-\frac12 k_L - k_L T \Big(\tilde{\omega}_L  - \frac{1}{4}\tilde{\mu} \Big)
+\frac{k_RT}{\tilde{\mu}} \Big( \pi^2 + \tilde{\omega}_R^2\Big)~, 
\end{eqnarray}
as in \eqref{eqn:lnZBPS}. 

Gibbs' free energy generates extensive variables through the first law of thermodynamics in the form
\be\la{eqn:1stlawG}
 dG  = - S dT  - P d\mu  -  J_L d\omega_L - J_R d\omega_R~.
\ee
Note that these are potentials without tilde, before rescaling by $T$.
For example, the entropy is given by a thermal derivative taken with fixed $\mu$, $\omega_L$, $\omega_R$:
\be\la{eqn:SfromG}
S= - \partial_T G  =  - \frac{2\pi^2 k_R}{\mu-1}  T + \frac{1}{4}k_L (2\pi)^2  T = S_*  + \pi^2 k_L T~,
\ee
up to linear order in $T$, in agreement with \eqref{eqn:SnearBPS}. In the nearBPS regime we can also quantify the magnitude of thermal {\it fluctuations} in the standard manner. For example, 
$$
\Big\langle j_L - \langle j_L  \rangle \Big\rangle^2 =  - \partial^2_{\omega_L} G = k_L~, 
$$
with the average value of $J_L = \langle j_L  \rangle = k_L$. The levels $k_{L,R}$ are both huge for semiclassical black holes, but they are finite. The relative fluctuations in the value of $J_L$ are of order $\sim k_L^{-\frac{1}{2}}$.


\section{Microscopics of the BPS Charge Constraint}\label{sec:micro}

In the previous sections, we reached the BPS limit of AdS$_3$ black holes from a thermodynamic point of view
and stressed that the supersymmetric limit is reached by tuning {\it two} potentials. 
In this section, we revisit this property of BPS black holes from a microscopic point of view, noting that
chiral primaries are co-dimension one in parameter space. 
We argue that black holes are ensemble averages, effectively restricting their macroscopic charges to
that of a particular chiral primary, thus yielding a second condition on the parameters. 
This gives a complementary and fully microscopic understanding of the BPS charge constraint $J_L = k_L$,
which was derived from the thermodynamic partition function in section \ref{sec:partfn}.

\subsection{Two-Dimensional Superconformal Algebra and Representations}

We consider black holes in AdS$_3 \times S^3$ described by supersymmetric CFT$_2$'s 
with $(4,4)$ supersymmetry. The (super-)conformal algebra simplifies greatly in two dimensions as it factorizes
into two independent copies of (super-)Virasoro algebra. 
To take advantage we first review the unitary representations of the small $\mathcal{N}=4$
superconformal algebra \cite{Sevrin:1988ew} in 2D \cite{Eguchi:1987sm,Eguchi:1987wf}.

It is sufficient to analyze one chiral sector of the $(4,4)$ algebra, either left or right, and we denote 
by $c=6k$ the central charge of this sector. Each unitary representation of the algebra is labeled by the $L_0$- and $J$-eigenvalues $(h,j)$
\footnote{We use the Dynkin convention where the label $j$ is always an integer and the $j$'th representation has dimension 
$j+1$. The half-integral spin familiar from quantum mechanics is $j_{\rm QM} =\frac{1}{2} j_{\rm here}$.}
of its superconformal primary, and the whole multiplet consists of the primary and its descendants.
We can focus on the NS sector because representations in the Ramond sector are 
\emph{isomorphic} through spectral flow by half-integral unit.
Then there are just two types of representations:  
the \emph{massless} (a.k.a. short) with a superconformal primary that saturates the unitarity bound
$h \geq \frac{1}{2}j$, and the \emph{massive} (a.k.a. long)  with a primary that does not.
The massless multiplets are enumerated by the representation of the $SU(2)$ R-charge in the range $j=0,1,\cdots,k$ that fixes 
the conformal weight $h=\frac{1}{2}j$.
The massive multiplets only permit the range $j=0,1, \cdots k-1$ but $h$ can take any real value strictly larger than the bound $h>\frac{1}{2}j$. 
Massive representations with identical $j$ and distinct $h$ all have the same structure so it is not
of our interest to distinguish them, they are not \emph{essentially} distinct.

The representations are conveniently described by their characters ${\rm Tr} ~q^h y^{j}$. Note that
the Casimir term $-\frac{c}{24}$ in the exponent is absent by convention, it must be restored in physical partition functions.
The character formulae for the two classes of multiplets are \cite{Eguchi:1987wf}:
\footnote{We turn off a $U(1)$ fugacity called $y$ in \cite{Eguchi:1987wf} 
and the $SU(2)$ fugacity is renamed $(e^{\frac12 i \theta})_{\rm there} = y_{\rm here}$.}
\bea\la{eqn:N4char}
\text{Massive : } ~{\rm ch}_{h,j}(q,y) &=& q^h F^{NS} 
\sum_{m=-\infty}^{\infty}  \left( y^{2(k+1)m +j+1}-y^{-2(k+1)m -j-1} \right) 
\frac{q^{(k+1)m^2+(j+1)m}}{y-y^{-1}} ~, \cr
(j=0,1, \cdots k-1) &&   \nonumber \\
\text{Massless : } ~~~\chi_{j}(q,y) &=& q^{\frac{j}{2}} F^{NS}
\sum_{m=-\infty}^{\infty}  \left( \frac{y^{2(k+1)m +j+1}}{(1+yq^{m+\frac12})^2}- \frac{y^{-2(k+1)m -j-1}}{(1+y^{-1}q^{m+\frac12})^2} \right)
\frac{q^{(k+1)m^2+(j+1)m}}{y-y^{-1}} ~, \cr
(j=0,1, \cdots k)  && 
\eea
where 
$$
F^{NS} = \prod_{n\geq1} \dfrac{\left( 1+yq^{n-\frac12} \right)^2 \left( 1+y^{-1}q^{n-\frac12} \right)^2 }{(1-y^2 q^n)(1-q^n)^2(1-y^{-2} q^n)} ~,
$$
accounts for the action of creation operators, i.e. the negative frequency modes $\{ G_{r<0}\}$ and $\{ L_{n<0}, J^i_{n<0}\}$ of the
four fermionic and four bosonic fields. Since the massive character ${\rm ch}_{h,j}(q,y)$ depends on the conformal weight $h$ only via $q^h$, it is convenient to define an 
$h$-independent massive character by shifting out the $h$ 
\emph{in excess of} the unitarity bound $\frac{1}{2}j$:
\be
\label{eqn:charshift}
\wt{\rm ch}_{j}(q,y) \equiv {\rm ch}_{h,j}(q,y) q^{-h+\frac{1}{2}j}~.
\ee

The transformation under spectral flow follows from these formulae. In particular, the sum over $m$ in \eqref{eqn:N4char} 
guarantees invariance of each character under spectral flow by integral $\eta$:
\begin{equation}
\begin{cases} h \to h_\eta = h - \eta j + k \eta^2 \\ j \to j_\eta = j - 2k\eta \end{cases}
\quad \Leftrightarrow \quad
q^h y^j \to q^{h_\eta} y^{j_\eta} = q^h (yq^{-\eta})^j q^{k \eta^2} y^{-2k\eta} ~.
\end{equation}
%

%
%
%

Although massless multiplets have no continuous parameter, it is possible that a combination of them
continuously deform into a massive multiplet, at least group theoretically.
Such \emph{recombination rules} are fairly simple.
Notice that the massless character formula  in \eqref{eqn:N4char} differ from the massive one only by the factors
$(1+y^{\pm 1} q^{m+\frac12})^2$ in the denominator.
Inspecting how $\chi_{j}(q,y)$ depends on $j$, one can see that these factors are
precisely cancelled by adding four characters with different $j$'s, thus yielding the mathematical identity:
\begin{equation}
\label{eqn:chchi}
\wt{\rm ch}_{j}(q,y) = \chi_{j}(q,y) + 2\chi_{j+1}(q,y) + \chi_{j+2}(q,y)~. 
\end{equation}
The identity holds literally for $j=0,\ldots,k-2$; for $j=k-1$ the term with index $j+2$ is undefined but the identity is valid with this
term omitted.

%
%
%
%
%
%

The supersymmetric index is protected against recombinations because contributions from the
four massless representations on the right hand side of  \eqref{eqn:chchi} cancel one another in the index, in agreement with the vanishing result for
the index of the massive representations with any value of $h$. The BPS partition function includes all massless representations and is not protected in this way. 

\subsection{Ensemble Average Gives the Charge Constraint}

Given the unitary representations described by their characters,
we are now ready to extract an extra constraint on macroscopic charges imposed by supersymmetry.

In the $\mathcal{N}=4$ superconformal algebra, the R-symmetry is $SU(2)$, rather than $SO(2)$ as in $\mathcal{N}=2$, 
so any chiral primary with $J$-eigenvalue $j$ is part of an $SU(2)$ representation that, 
in particular, contains the anti-chiral primary with $J$-eigenvalue $-j$ and
the same $L_0$-eigenvalue $h=\frac{1}{2}j$ as the initial chiral primary,
which saturates the anti-chiral unitarity bound $h \geq -\frac{1}{2}j$.
The anti-chiral primary is related to a state with eigenvalues $(h,j) = (k-\frac{1}{2}j, 2k-j)$ via spectral flow and this state is itself 
a chiral primary. Thus a chiral primary with R-charge $j$ always comes in pair with another chiral primary that has R-charge $2k-j$.

This pairing is easily observed in explicit expansion of the characters (\ref{eqn:N4char}). For example, for $k=5$,
\begin{eqnarray}
\label{eqn:charex}
\chi_{j=0}(q,y) &=& 1 + y^2q + y^4q^2 + y^6q^3 + y^8q^4 +y^{10}q^5 +   \cdots ~, \nonumber \\
\chi_{j=1}(q,y) &=& yq^{1/2} + y^3q^{3/2} + y^5q^{5/2} + y^7q^{7/2} + y^9q^{9/2} +   \cdots ~, \nonumber
\end{eqnarray}
where ellipses represent terms with strictly $h > \frac{1}{2}j$.

In fact the argument is not restricted to chiral primaries, it shows that
any state with eigenvalues $(h,j)$ is paired with another 
state with $(h+k-j, 2k-j)$.
The pair is characterized by having R-charge mirrored about $k$ 
and the same conformal weight \emph{in excess of} the unitarity bound:
$$
h - \frac{1}{2}j = (h+k-j) - \frac12 (2k-j)~.
$$
The claim can be explicitly proved using the characters (\ref{eqn:N4char}). The
${\mathbb Z}_2$ exchange operation within the pairs corresponds to 
a substitution $y \to q^{-1} y^{-1}$ followed by multiplication by $q^k y^{2k}$, because
\begin{equation}
\label{eqn:swap}
q^h y^j \to q^h (q^{-1}y^{-1})^{j} q^k y^{2k} = q^{h+k-j} y^{2k-j} ~.
\end{equation}
Then one can verify that all characters (\ref{eqn:N4char}) are invariant (or, \emph{even}) 
under this ${\mathbb Z}_2$ transformation:
\bea\la{eqn:Z2even}
\chi_{j}(q,y) &=& \chi_{j}(q,q^{-1}y^{-1}) \cdot  q^k y^{2k} ~, \nonumber \\
\wt{\rm ch}_{j}(q,y) &=& \wt{\rm ch}_{j}(q,q^{-1}y^{-1}) \cdot  q^k y^{2k} ~,
\eea
proving that all states appear in pairs, as claimed.

Provided an ensemble of microscopic states that come packaged in multiplets,
macroscopic charges are obtained by taking ensemble averages.
We have seen that every state within any multiplet comes in a pair with another state
with respective R-charges $j$ and $2k-j$.
It is obvious that the ensemble average of the angular momentum turns out to be $k$,
regardless of which and how many multiplets of each type appear in the ensemble.
 
An important caveat in this argument is that both microscopic states within a pair
must be weighed with equal probability within the canonical ensemble.
Given the eigenvalues $(h,j)$ and $(h+k-j, 2k-j)$ of the two states,
this assumption translates into a relation between chemical potentials:
\begin{equation}
\label{eqn:equalprob}
\tau + 2z=0 ~,
\end{equation}
where $\tau$ and $z$ define the canonical partition function by
\be\la{eqn:canpf}
Z = {\rm Tr}~ e^{2 \pi i \tau L_0 + 2 \pi i z J} ~.
\ee

To see how this argument applies to microscopic accounting of BPS black holes,
we start again from the definition of the partition function \eqref{eqn:partdef},
as rewritten in \eqref{eqn:ZBPS}:
\bea\la{eqn:partdefrep}
Z 
&=& e^{\frac{1}{2}\beta k_L} {\rm Tr} ~e^{-\beta(\epsilon -p - j_L + \frac{1}{2}k_L)    + \tilde{\mu} p + \tilde{\omega}_R j_R  +\tilde{\omega}_L  j_L} ~,
\eea
where we recall the definitions $\tilde\mu = \beta (\mu-1)$, $\tilde\omega_R = \beta\omega_R$, and $\tilde\omega_L = \beta (\omega_L-1)$.
Our interest is the pairing in the $L$-sector where for a state with quantum numbers $(p,j_L, j_R)$
that saturates the BPS bound there is another BPS state that has quantum numbers $(p-k_L+ j_L, 2k_L-j_L ,j_R)$.
In the supersymmetric partition function we choose potentials so
\be\la{eqn:potcons}
-\tilde\mu + 2\tilde\omega_L =0 ~,
\ee
which guarantees that the two members of the pair have the same weight.
It follows that the contribution from the two states in the pair to the expectation value $\langle j_L\rangle$ is $k_L$. 

The discussion in this subsection is based on the partition function and we do not appeal to cancellations, 
unlike in the reasoning based on the index.
Rather, we interpret the splitting and joining of the BPS states in the chiral ring as a thermodynamic process
where there are many possible values of the quantum number $j_L$ but, in the ensemble realized by a black hole,
thermodynamic equilibrium forces the macroscopic value $J_L = k_L$,
even though this is not the value in most microstates by themselves.

\section*{Acknowledgements}

We would like to thank James Liu, Jun Nian, Leopoldo Pando Zayas, Shruti Paranjape and Yangwenxiao Zeng for helpful discussions and communications.
This work was supported in part by the U.S. Department of Energy under grant DE-SC0007859.

\bibliography{AdS3draftFLFinal}{}

\providecommand{\href}[2]{#2}\begingroup\raggedright\begin{thebibliography}{10}

\bibitem{Strominger:1996sh}
A.~Strominger and C.~Vafa, \emph{{Microscopic origin of the Bekenstein-Hawking
  entropy}}, \href{https://doi.org/10.1016/0370-2693(96)00345-0}{\emph{Phys.
  Lett. B} {\bfseries 379} (1996) 99}
  [\href{https://arxiv.org/abs/hep-th/9601029}{{\ttfamily hep-th/9601029}}].

\bibitem{Maldacena:1997re}
J.~M. Maldacena, \emph{{The Large N limit of superconformal field theories and
  supergravity}}, \href{https://doi.org/10.1023/A:1026654312961}{\emph{Int. J.
  Theor. Phys.} {\bfseries 38} (1999) 1113}
  [\href{https://arxiv.org/abs/hep-th/9711200}{{\ttfamily hep-th/9711200}}].

\bibitem{Aharony:2003sx}
O.~Aharony, J.~Marsano, S.~Minwalla, K.~Papadodimas and M.~Van~Raamsdonk,
  \emph{{The Hagedorn - deconfinement phase transition in weakly coupled large
  N gauge theories}},
  \href{https://doi.org/10.4310/ATMP.2004.v8.n4.a1}{\emph{Adv. Theor. Math.
  Phys.} {\bfseries 8} (2004) 603}
  [\href{https://arxiv.org/abs/hep-th/0310285}{{\ttfamily hep-th/0310285}}].

\bibitem{Kinney:2005ej}
J.~Kinney, J.~M. Maldacena, S.~Minwalla and S.~Raju, \emph{{An Index for 4
  dimensional super conformal theories}},
  \href{https://doi.org/10.1007/s00220-007-0258-7}{\emph{Commun. Math. Phys.}
  {\bfseries 275} (2007) 209}
  [\href{https://arxiv.org/abs/hep-th/0510251}{{\ttfamily hep-th/0510251}}].

\bibitem{Romelsberger:2005eg}
C.~Romelsberger, \emph{{Counting chiral primaries in N = 1, d=4 superconformal
  field theories}},
  \href{https://doi.org/10.1016/j.nuclphysb.2006.03.037}{\emph{Nucl. Phys. B}
  {\bfseries 747} (2006) 329}
  [\href{https://arxiv.org/abs/hep-th/0510060}{{\ttfamily hep-th/0510060}}].

\bibitem{Berkooz:2008gc}
M.~Berkooz and D.~Reichmann, \emph{{Weakly Renormalized Near 1/16 SUSY Fermi
  Liquid Operators in N=4 SYM}},
  \href{https://doi.org/10.1088/1126-6708/2008/10/084}{\emph{JHEP} {\bfseries
  10} (2008) 084} [\href{https://arxiv.org/abs/0807.0559}{{\ttfamily
  0807.0559}}].

\bibitem{Chang:2013fba}
C.-M. Chang and X.~Yin, \emph{{1/16 BPS states in $\mathcal N=$ 4
  super-Yang-Mills theory}},
  \href{https://doi.org/10.1103/PhysRevD.88.106005}{\emph{Phys. Rev. D}
  {\bfseries 88} (2013) 106005}
  [\href{https://arxiv.org/abs/1305.6314}{{\ttfamily 1305.6314}}].

\bibitem{Benini:2015noa}
F.~Benini and A.~Zaffaroni, \emph{{A topologically twisted index for
  three-dimensional supersymmetric theories}},
  \href{https://doi.org/10.1007/JHEP07(2015)127}{\emph{JHEP} {\bfseries 07}
  (2015) 127} [\href{https://arxiv.org/abs/1504.03698}{{\ttfamily
  1504.03698}}].

\bibitem{Benini:2015eyy}
F.~Benini, K.~Hristov and A.~Zaffaroni, \emph{{Black hole microstates in
  AdS$_{4}$ from supersymmetric localization}},
  \href{https://doi.org/10.1007/JHEP05(2016)054}{\emph{JHEP} {\bfseries 05}
  (2016) 054} [\href{https://arxiv.org/abs/1511.04085}{{\ttfamily
  1511.04085}}].

\bibitem{Benini:2016rke}
F.~Benini, K.~Hristov and A.~Zaffaroni, \emph{{Exact microstate counting for
  dyonic black holes in AdS4}},
  \href{https://doi.org/10.1016/j.physletb.2017.05.076}{\emph{Phys. Lett. B}
  {\bfseries 771} (2017) 462}
  [\href{https://arxiv.org/abs/1608.07294}{{\ttfamily 1608.07294}}].

\bibitem{Hosseini:2017mds}
S.~M. Hosseini, K.~Hristov and A.~Zaffaroni, \emph{{An extremization principle
  for the entropy of rotating BPS black holes in AdS$_{5}$}},
  \href{https://doi.org/10.1007/JHEP07(2017)106}{\emph{JHEP} {\bfseries 07}
  (2017) 106} [\href{https://arxiv.org/abs/1705.05383}{{\ttfamily
  1705.05383}}].

\bibitem{Cabo-Bizet:2018ehj}
A.~Cabo-Bizet, D.~Cassani, D.~Martelli and S.~Murthy, \emph{{Microscopic origin
  of the Bekenstein-Hawking entropy of supersymmetric AdS$_{5}$ black holes}},
  \href{https://doi.org/10.1007/JHEP10(2019)062}{\emph{JHEP} {\bfseries 10}
  (2019) 062} [\href{https://arxiv.org/abs/1810.11442}{{\ttfamily
  1810.11442}}].

\bibitem{Choi:2018hmj}
S.~Choi, J.~Kim, S.~Kim and J.~Nahmgoong, \emph{{Large AdS black holes from
  QFT}},  \href{https://arxiv.org/abs/1810.12067}{{\ttfamily 1810.12067}}.

\bibitem{Benini:2018ywd}
F.~Benini and P.~Milan, \emph{{Black Holes in 4D $\mathcal{N}$=4
  Super-Yang-Mills Field Theory}},
  \href{https://doi.org/10.1103/PhysRevX.10.021037}{\emph{Phys. Rev. X}
  {\bfseries 10} (2020) 021037}
  [\href{https://arxiv.org/abs/1812.09613}{{\ttfamily 1812.09613}}].

\bibitem{Choi:2019miv}
S.~Choi and S.~Kim, \emph{{Large AdS$_6$ black holes from CFT$_5$}},
  \href{https://arxiv.org/abs/1904.01164}{{\ttfamily 1904.01164}}.

\bibitem{Zaffaroni:2019dhb}
A.~Zaffaroni, \emph{{Lectures on AdS Black Holes, Holography and
  Localization}},
  \href{https://doi.org/10.1007/s41114-020-00027-8}{\emph{Living Rev. Rel.}
  {\bfseries 23} (2020) 2} [\href{https://arxiv.org/abs/1902.07176}{{\ttfamily
  1902.07176}}].

\bibitem{Gutowski:2004ez}
J.~B. Gutowski and H.~S. Reall, \emph{{Supersymmetric AdS(5) black holes}},
  \href{https://doi.org/10.1088/1126-6708/2004/02/006}{\emph{JHEP} {\bfseries
  02} (2004) 006} [\href{https://arxiv.org/abs/hep-th/0401042}{{\ttfamily
  hep-th/0401042}}].

\bibitem{Gutowski:2004yv}
J.~B. Gutowski and H.~S. Reall, \emph{{General supersymmetric AdS(5) black
  holes}}, \href{https://doi.org/10.1088/1126-6708/2004/04/048}{\emph{JHEP}
  {\bfseries 04} (2004) 048}
  [\href{https://arxiv.org/abs/hep-th/0401129}{{\ttfamily hep-th/0401129}}].

\bibitem{Chong:2005da}
Z.~Chong, M.~Cvetic, H.~Lu and C.~Pope, \emph{{Five-dimensional gauged
  supergravity black holes with independent rotation parameters}},
  \href{https://doi.org/10.1103/PhysRevD.72.041901}{\emph{Phys. Rev. D}
  {\bfseries 72} (2005) 041901}
  [\href{https://arxiv.org/abs/hep-th/0505112}{{\ttfamily hep-th/0505112}}].

\bibitem{Chong:2005hr}
Z.-W. Chong, M.~Cvetic, H.~Lu and C.~Pope, \emph{{General non-extremal rotating
  black holes in minimal five-dimensional gauged supergravity}},
  \href{https://doi.org/10.1103/PhysRevLett.95.161301}{\emph{Phys. Rev. Lett.}
  {\bfseries 95} (2005) 161301}
  [\href{https://arxiv.org/abs/hep-th/0506029}{{\ttfamily hep-th/0506029}}].

\bibitem{Kunduri:2006ek}
H.~K. Kunduri, J.~Lucietti and H.~S. Reall, \emph{{Supersymmetric multi-charge
  AdS(5) black holes}},
  \href{https://doi.org/10.1088/1126-6708/2006/04/036}{\emph{JHEP} {\bfseries
  04} (2006) 036} [\href{https://arxiv.org/abs/hep-th/0601156}{{\ttfamily
  hep-th/0601156}}].

\bibitem{Wu:2011gq}
S.-Q. Wu, \emph{{General Nonextremal Rotating Charged AdS Black Holes in
  Five-dimensional $U(1)^3$ Gauged Supergravity: A Simple Construction
  Method}}, \href{https://doi.org/10.1016/j.physletb.2011.12.031}{\emph{Phys.
  Lett. B} {\bfseries 707} (2012) 286}
  [\href{https://arxiv.org/abs/1108.4159}{{\ttfamily 1108.4159}}].

\bibitem{Kim:2006he}
S.~Kim and K.-M. Lee, \emph{{1/16-BPS Black Holes and Giant Gravitons in the
  AdS(5) X S**5 Space}},
  \href{https://doi.org/10.1088/1126-6708/2006/12/077}{\emph{JHEP} {\bfseries
  12} (2006) 077} [\href{https://arxiv.org/abs/hep-th/0607085}{{\ttfamily
  hep-th/0607085}}].

\bibitem{Kim:2019yrz}
J.~Kim, S.~Kim and J.~Song, \emph{{A 4d N=1 Cardy Formula}},
  \href{https://arxiv.org/abs/1904.03455}{{\ttfamily 1904.03455}}.

\bibitem{Cabo-Bizet:2020nkr}
A.~Cabo-Bizet, D.~Cassani, D.~Martelli and S.~Murthy, \emph{{The large-$N$
  limit of the 4d $ \mathcal{N} $ = 1 superconformal index}},
  \href{https://doi.org/10.1007/JHEP11(2020)150}{\emph{JHEP} {\bfseries 11}
  (2020) 150} [\href{https://arxiv.org/abs/2005.10654}{{\ttfamily
  2005.10654}}].

\bibitem{Murthy:2020rbd}
S.~Murthy, \emph{{The growth of the $\frac{1}{16}$-BPS index in 4d
  $\mathcal{N}=4$ SYM}},  \href{https://arxiv.org/abs/2005.10843}{{\ttfamily
  2005.10843}}.

\bibitem{Agarwal:2020zwm}
P.~Agarwal, S.~Choi, J.~Kim, S.~Kim and J.~Nahmgoong, \emph{{AdS black holes
  and finite N indices}},  \href{https://arxiv.org/abs/2005.11240}{{\ttfamily
  2005.11240}}.

\bibitem{GonzalezLezcano:2020yeb}
A.~Gonz\'alez~Lezcano, J.~Hong, J.~T. Liu and L.~A. Pando~Zayas,
  \emph{{Sub-leading Structures in Superconformal Indices: Subdominant Saddles
  and Logarithmic Contributions}},
  \href{https://arxiv.org/abs/2007.12604}{{\ttfamily 2007.12604}}.

\bibitem{Copetti:2020dil}
C.~Copetti, A.~Grassi, Z.~Komargodski and L.~Tizzano, \emph{{Delayed
  Deconfinement and the Hawking-Page Transition}},
  \href{https://arxiv.org/abs/2008.04950}{{\ttfamily 2008.04950}}.

\bibitem{Goldstein:2020yvj}
K.~Goldstein, V.~Jejjala, Y.~Lei, S.~van Leuven and W.~Li, \emph{{Residues,
  modularity, and the Cardy limit of the 4d $\mathcal{N}=4$ superconformal
  index}},  \href{https://arxiv.org/abs/2011.06605}{{\ttfamily 2011.06605}}.

\bibitem{Larsen:2019oll}
F.~Larsen, J.~Nian and Y.~Zeng, \emph{{AdS$_{5}$ black hole entropy near the
  BPS limit}}, \href{https://doi.org/10.1007/JHEP06(2020)001}{\emph{JHEP}
  {\bfseries 06} (2020) 001}
  [\href{https://arxiv.org/abs/1907.02505}{{\ttfamily 1907.02505}}].

\bibitem{Razamat:2012uv}
S.~S. Razamat, \emph{{On a modular property of N=2 superconformal theories in
  four dimensions}}, \href{https://doi.org/10.1007/JHEP10(2012)191}{\emph{JHEP}
  {\bfseries 10} (2012) 191} [\href{https://arxiv.org/abs/1208.5056}{{\ttfamily
  1208.5056}}].

\bibitem{Gadde:2020bov}
A.~Gadde, \emph{{Modularity of supersymmetric partition functions}},
  \href{https://arxiv.org/abs/2004.13490}{{\ttfamily 2004.13490}}.

\bibitem{Eguchi:1987sm}
T.~Eguchi and A.~Taormina, \emph{{Unitary Representations of $N=4$
  Superconformal Algebra}},
  \href{https://doi.org/10.1016/0370-2693(87)91679-0}{\emph{Phys. Lett. B}
  {\bfseries 196} (1987) 75}.

\bibitem{Eguchi:1987wf}
T.~Eguchi and A.~Taormina, \emph{{Character Formulas for the $N=4$
  Superconformal Algebra}},
  \href{https://doi.org/10.1016/0370-2693(88)90778-2}{\emph{Phys. Lett. B}
  {\bfseries 200} (1988) 315}.

\bibitem{Kraus:2006nb}
P.~Kraus and F.~Larsen, \emph{{Partition functions and elliptic genera from
  supergravity}},
  \href{https://doi.org/10.1088/1126-6708/2007/01/002}{\emph{JHEP} {\bfseries
  01} (2007) 002} [\href{https://arxiv.org/abs/hep-th/0607138}{{\ttfamily
  hep-th/0607138}}].

\bibitem{Cvetic:1998xh}
M.~Cvetic and F.~Larsen, \emph{{Near horizon geometry of rotating black holes
  in five-dimensions}},
  \href{https://doi.org/10.1016/S0550-3213(98)00604-X}{\emph{Nucl. Phys. B}
  {\bfseries 531} (1998) 239}
  [\href{https://arxiv.org/abs/hep-th/9805097}{{\ttfamily hep-th/9805097}}].

\bibitem{Gimon:2007mh}
E.~G. Gimon, F.~Larsen and J.~Simon, \emph{{Black holes in Supergravity: The
  Non-BPS branch}},
  \href{https://doi.org/10.1088/1126-6708/2008/01/040}{\emph{JHEP} {\bfseries
  01} (2008) 040} [\href{https://arxiv.org/abs/0710.4967}{{\ttfamily
  0710.4967}}].

\bibitem{Assel:2015nca}
B.~Assel, D.~Cassani, L.~Di~Pietro, Z.~Komargodski, J.~Lorenzen and
  D.~Martelli, \emph{{The Casimir Energy in Curved Space and its Supersymmetric
  Counterpart}}, \href{https://doi.org/10.1007/JHEP07(2015)043}{\emph{JHEP}
  {\bfseries 07} (2015) 043}
  [\href{https://arxiv.org/abs/1503.05537}{{\ttfamily 1503.05537}}].

\bibitem{Larsen:2020lhg}
F.~Larsen and S.~Paranjape, \emph{{Thermodynamics of Near BPS Black Holes in
  AdS$_4$ and AdS$_7$}},  \href{https://arxiv.org/abs/2010.04359}{{\ttfamily
  2010.04359}}.

\bibitem{Larsen:2018iou}
F.~Larsen, \emph{{A nAttractor mechanism for nAdS$_{2}$/nCFT$_{1}$
  holography}}, \href{https://doi.org/10.1007/JHEP04(2019)055}{\emph{JHEP}
  {\bfseries 04} (2019) 055}
  [\href{https://arxiv.org/abs/1806.06330}{{\ttfamily 1806.06330}}].

\bibitem{Hong:2019tsx}
J.~Hong, F.~Larsen and J.~T. Liu, \emph{{The scales of black holes with
  nAdS$_{2}$ geometry}},
  \href{https://doi.org/10.1007/JHEP10(2019)260}{\emph{JHEP} {\bfseries 10}
  (2019) 260} [\href{https://arxiv.org/abs/1907.08862}{{\ttfamily
  1907.08862}}].

\bibitem{Banados:1992wn}
M.~Banados, C.~Teitelboim and J.~Zanelli, \emph{{The Black hole in
  three-dimensional space-time}},
  \href{https://doi.org/10.1103/PhysRevLett.69.1849}{\emph{Phys. Rev. Lett.}
  {\bfseries 69} (1992) 1849}
  [\href{https://arxiv.org/abs/hep-th/9204099}{{\ttfamily hep-th/9204099}}].

\bibitem{Banados:1992gq}
M.~Banados, M.~Henneaux, C.~Teitelboim and J.~Zanelli, \emph{{Geometry of the
  (2+1) black hole}},
  \href{https://doi.org/10.1103/PhysRevD.48.1506}{\emph{Phys. Rev. D}
  {\bfseries 48} (1993) 1506}
  [\href{https://arxiv.org/abs/gr-qc/9302012}{{\ttfamily gr-qc/9302012}}].

\bibitem{Brown:1986nw}
J.~Brown and M.~Henneaux, \emph{{Central Charges in the Canonical Realization
  of Asymptotic Symmetries: An Example from Three-Dimensional Gravity}},
  \href{https://doi.org/10.1007/BF01211590}{\emph{Commun. Math. Phys.}
  {\bfseries 104} (1986) 207}.

\bibitem{Kraus:2005zm}
P.~Kraus and F.~Larsen, \emph{{Holographic gravitational anomalies}},
  \href{https://doi.org/10.1088/1126-6708/2006/01/022}{\emph{JHEP} {\bfseries
  01} (2006) 022} [\href{https://arxiv.org/abs/hep-th/0508218}{{\ttfamily
  hep-th/0508218}}].

\bibitem{Witten:1998zw}
E.~Witten, \emph{{Anti-de Sitter space, thermal phase transition, and
  confinement in gauge theories}},
  \href{https://doi.org/10.4310/ATMP.1998.v2.n3.a3}{\emph{Adv. Theor. Math.
  Phys.} {\bfseries 2} (1998) 505}
  [\href{https://arxiv.org/abs/hep-th/9803131}{{\ttfamily hep-th/9803131}}].

\bibitem{Choi:2018vbz}
S.~Choi, J.~Kim, S.~Kim and J.~Nahmgoong, \emph{{Comments on deconfinement in
  AdS/CFT}},  \href{https://arxiv.org/abs/1811.08646}{{\ttfamily 1811.08646}}.

\bibitem{Dijkgraaf:2000fq}
R.~Dijkgraaf, J.~M. Maldacena, G.~W. Moore and E.~P. Verlinde, \emph{{A Black
  hole Farey tail}},  \href{https://arxiv.org/abs/hep-th/0005003}{{\ttfamily
  hep-th/0005003}}.

\bibitem{Sevrin:1988ew}
A.~Sevrin, W.~Troost and A.~Van~Proeyen, \emph{{Superconformal Algebras in
  Two-Dimensions with N=4}},
  \href{https://doi.org/10.1016/0370-2693(88)90645-4}{\emph{Phys. Lett.}
  {\bfseries B208} (1988) 447}.

\end{thebibliography}\endgroup
\bibliographystyle{JHEP}

\end{document}